\RequirePackage{fix-cm}
\documentclass[smallextended]{svjour3}       
\smartqed  
\usepackage{graphicx}
\usepackage{bm}
\usepackage{url}
\usepackage{multirow}
\usepackage{tikz-cd}
\usepackage{tikz}
\usepackage{amsmath}
\usepackage{amssymb}
\usepackage{algorithm}
\usepackage{algorithmic}
\usepackage{caption}
\usepackage{footmisc}
\usepackage{subfigure}
\usepackage{natbib}
\usetikzlibrary{positioning}
\usetikzlibrary{shapes,snakes}

\begin{document}

\title{Attributed Network Embedding via Subspace Discovery
}


\author{Daokun Zhang         \and
        Jie Yin \and\\
        Xingquan Zhu \and
        Chengqi Zhang
}


\institute{Daokun Zhang \at
              Centre for Artificial Intelligence, FEIT, University of Technology Sydney, Australia \\
              \email{Daokun.Zhang@student.uts.edu.au} 
           \and
           Jie Yin
            \at
            Discipline of Business Analytics,
            The University of Sydney, Australia\\
            \email{jie.yin@sydney.edu.au}
            \and
            Xingquan Zhu \at
            Department of CEECS, 
            Florida Atlantic University, USA\\
            \email{xqzhu@cse.fau.edu}
            \and
            Chengqi Zhang \at
            Centre for Artificial Intelligence, FEIT, University of Technology Sydney, Australia \\
            \email{Chengqi.Zhang@uts.edu.au} 
}

\date{Received: date / Accepted: date}

\maketitle

\begin{abstract}
Network embedding aims to learn a latent, low-dimensional vector representations of network nodes, effective in supporting various network analytic tasks. While prior arts on network embedding focus primarily on preserving network topology structure to learn node representations, recently proposed attributed network embedding algorithms attempt to integrate rich node content information with network topological structure for enhancing the quality of network embedding. In reality, networks often have sparse content, incomplete node attributes, as well as the discrepancy between node attribute feature space and network structure space, which severely deteriorates the performance of existing methods. In this paper, we propose a unified framework for attributed network embedding--attri2vec--that learns node embeddings by discovering a latent node attribute subspace via a network structure guided transformation performed on the original attribute space. The resultant latent subspace can respect network structure in a more consistent way towards learning high-quality node representations. We formulate an optimization problem which is solved by an efficient stochastic gradient descent algorithm, with linear time complexity to the number of nodes. We investigate a series of linear and non-linear transformations performed on node attributes and empirically validate their effectiveness on various types of networks. Another advantage of attri2vec is its ability to solve out-of-sample problems, where embeddings of new coming nodes can be inferred from their node attributes through the learned mapping function. Experiments on various types of networks confirm that attri2vec is superior to state-of-the-art baselines for node classification, node clustering, as well as out-of-sample link prediction tasks. The source code of this paper is available at \url{https://github.com/daokunzhang/attri2vec}.
\end{abstract}


\section{Introduction}
\label{sec:intro}

With the ubiquity of network data across a diverse set of fields, network embedding has aroused a surge of research interests because it provides a revolutionized solution to network data analysis. Network embedding aims at learning lower-dimensional vector representations of network nodes, which allows downstream tasks, such as node classification, link prediction, or community detection, to be efficiently solved in the new vector space by simply applying traditional vector-based machine learning algorithms. The essential principle is to learn low-dimensional node representations that are able to preserve the proximity in the original network space, such that nodes with larger proximity would have similar vector representations. 

Traditionally, network embedding techniques (e.g., DeepWalk~\citep{perozzi2014deepwalk}, LINE~\citep{tang2015line}, node2vec~\citep{grover2016node2vec}, GraRep ~\citep{cao2015grarep}) mainly focus on preserving local or global network structure, making structurally similar nodes to be represented close to each other in the new vector space. In real-world scenarios, however, apart from structural information, network nodes are often associated with content, such as text features in citation networks, and user profiles in social networks. Node attributes not only provide direct evidence about content-level proximity but also influence the forming of interactions between nodes. It has been shown that, by jointly learning node embeddings with network structure and node attributes, attributed network embedding can achieve better performance. For example, TADW~\citep{yang2015network} exploits inductive matrix factorization~\citep{natarajan2014inductive} to incorporate textual features of nodes into the learning of node representations, showing better performance than DeepWalk. HSCA~\citep{zhang2016homophily} learns informative node representations through simultaneously integrating structural context with node content while enforcing the homophily property in the representation space. 

\begin{figure*}[t]
	\centering
	\subfigure[degree]{
		\label{fig:dist:subfig:degree}
		\includegraphics[width=2.25in]{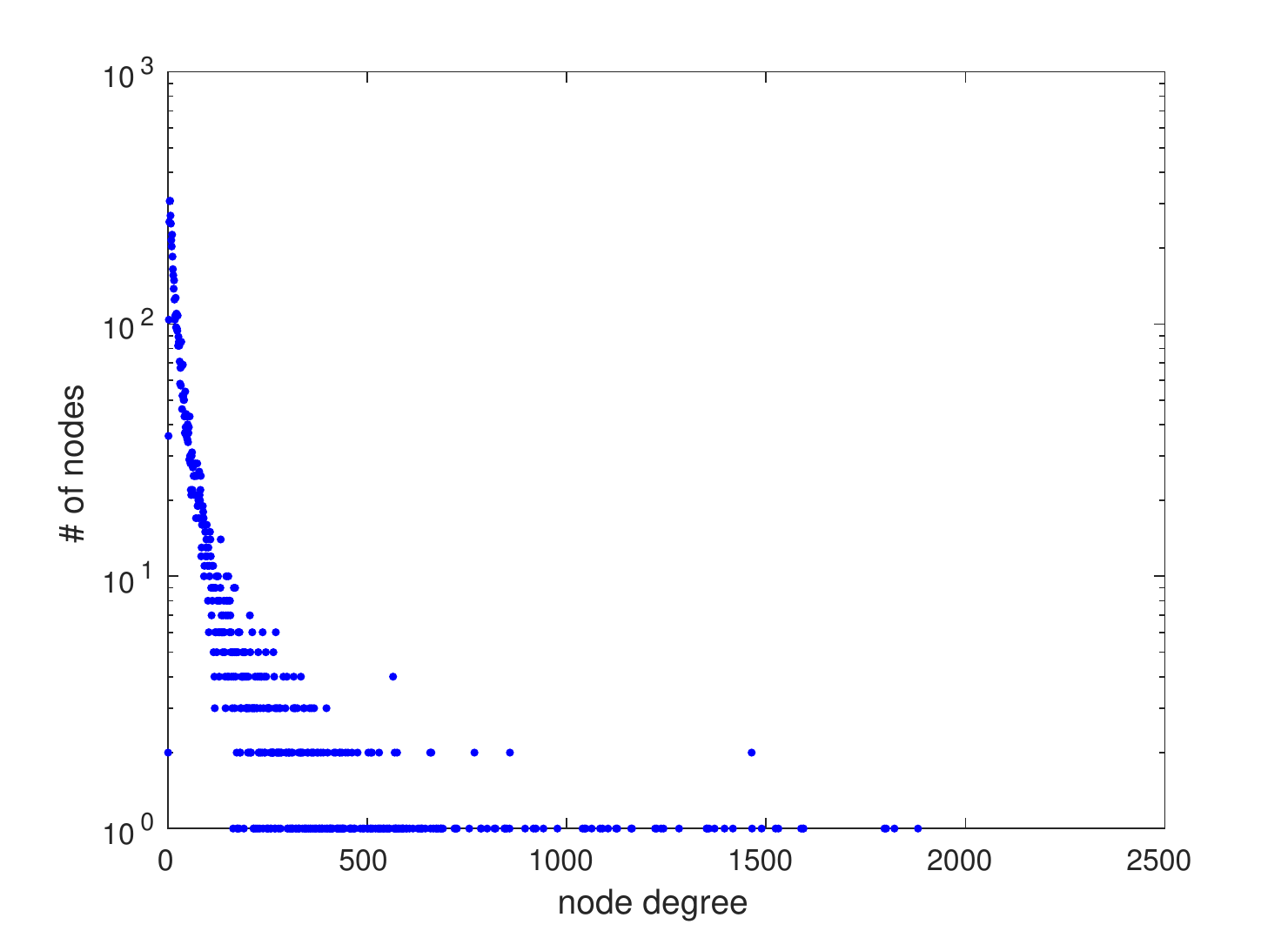}}
	\subfigure[attribute number]{
		\label{fig:dist:subfig:attribute}
		\includegraphics[width=2.25in]{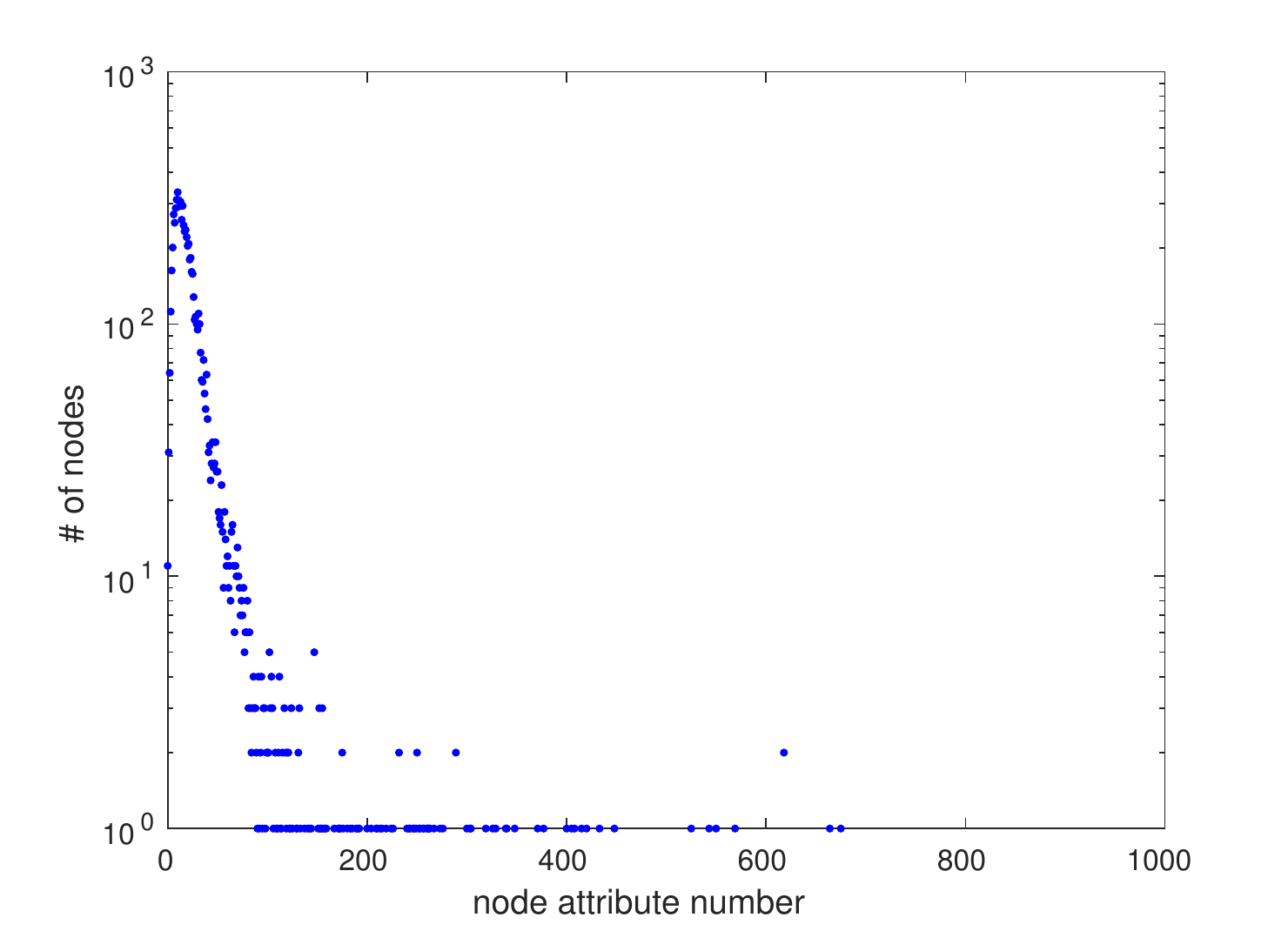}}
	\caption{Node distributions with respect to node degree and the number of node attributes on the Flickr network.}
	\label{fig:dist} 
\end{figure*}

Despite existing methods' demonstrated effectiveness in improving embedding performance, attributed network embedding is confronted with two major challenges. The first challenge lies in the fact that real-world networks are often very sparse; observable links connecting network nodes are rather limited, and node content information is sparse or incomplete. For example, in Figs.~\ref{fig:dist:subfig:degree} and \ref{fig:dist:subfig:attribute}, we plot the node distributions with respect to node degree and attribute number of a Flickr network, which contains 7,575 users and 12,047 user tag attributes. The vertical axis of Figs.~\ref{fig:dist:subfig:degree} and \ref{fig:dist:subfig:attribute} indicates the number of nodes, and the horizontal axis indicates node degree, and the number of node attributes, respectively. Figs.~\ref{fig:dist:subfig:degree} and \ref{fig:dist:subfig:attribute} show that the two plots both follow power-law distributions~\citep{guo2018cfond}, where only a small number of nodes have a high degree or a large number of node attributes. In other words, a significant number of nodes tend to have very few or missing links as well as incomplete attribute information. The sparsity in network structure and node attributes can negatively impact the estimation of structural-level and content-level proximity, thus further deteriorating the performance of network embedding.

\begin{figure*}[t]
	\centering
	\includegraphics[width=3in]{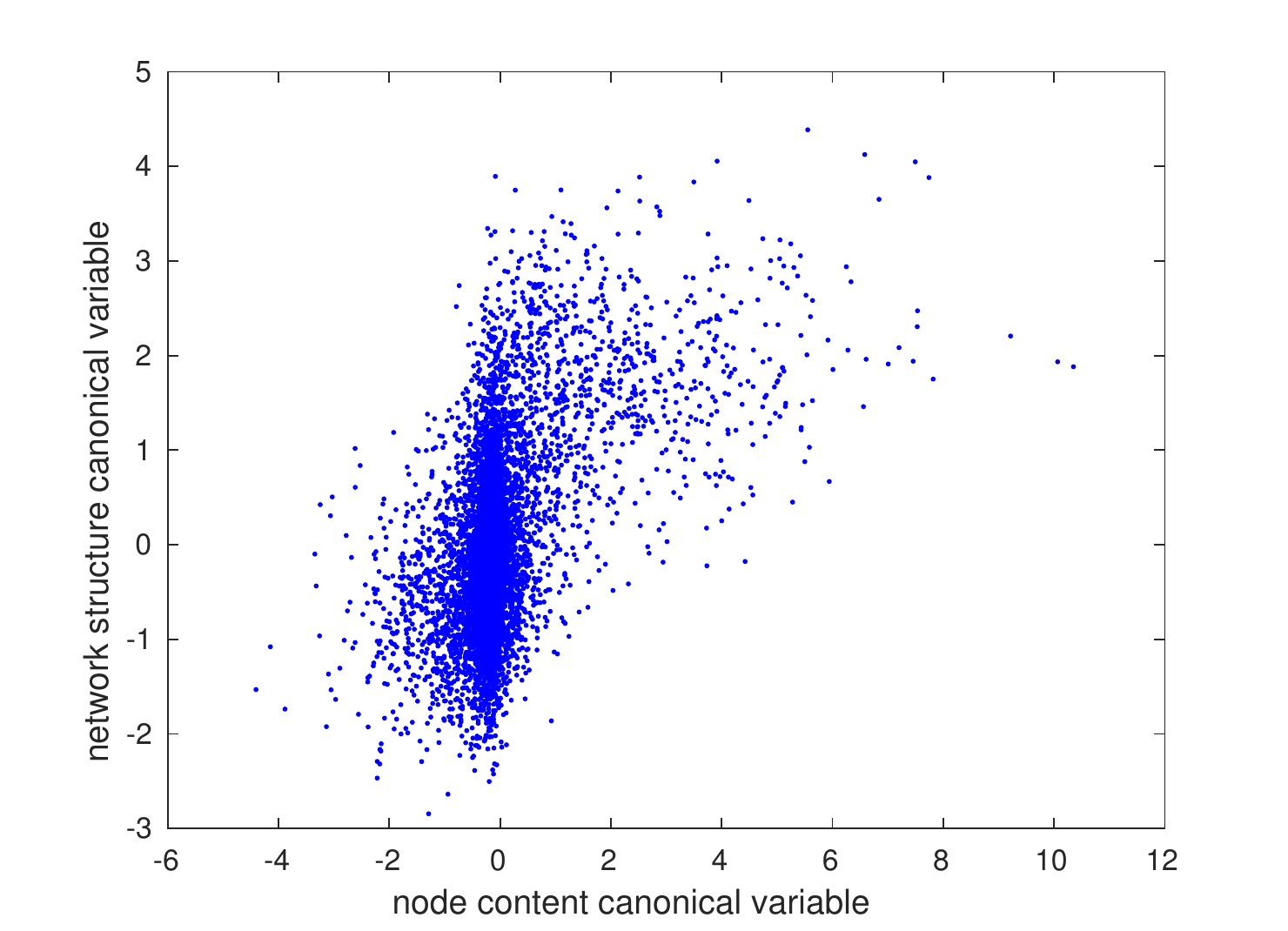}
	\caption{The scatter plot of node content canonical variable and network structure canonical variable.}
	\label{fig:canonical_scatter} 
\end{figure*}

The second challenge stems from the heterogeneity between network structure and node attributes. According to the social influence principle, network structure often exhibits certain correlations with node attributes~\citep{reagans2003network}. For example, in social networks, users who are connected are more likely to have similar attributes. However, because network structure and node attributes are two heterogeneous information sources, inevitably, there is a discrepancy between the two different feature spaces. As has been revealed in previous studies~\citep{Bianconi2009Assessing,subbarai2015what}, social network users with similar attributes, or antithetical attributes are both likely to connect to one another. To illustrate this point, let us consider a case study on the Flickr network, where we conduct correlation analysis between node representations learned from network structure and those from node attributes. Specifically, we obtain two sets of 128-dimensional node representations by applying DeepWalk~\citep{perozzi2014deepwalk} on network structure, and by performing Singular Value Decomposition (SVD) on node content features, and then conduct Canonical Correlation Analysis (CCA)~\citep{hotelling1936relations} on the two sets of generated representations. Fig.~\ref{fig:canonical_scatter} shows the scatter plot about node content canonical variable and network structure canonical variable, which are the linear combination of 128-dimensional structure representation and attribute representation, respectively, with maximum correlation. The correlation between the two canonical variables is 0.525. As is shown in the figure, most content canonical variables concentrate around 0, while most network structure canonical variables span from -2 to 2, which proves that the two canonical variables are not well correlated. The CCA results show that network structure and node content features do not always exhibit strong linear correlations. Such a discrepancy adds extra difficulties in fusing node attributes with network structure in a proper way that they complement rather than deteriorate each other towards learning high-quality node representations.

\begin{figure*}[t]
	\centering
	\includegraphics[width=3.6in]{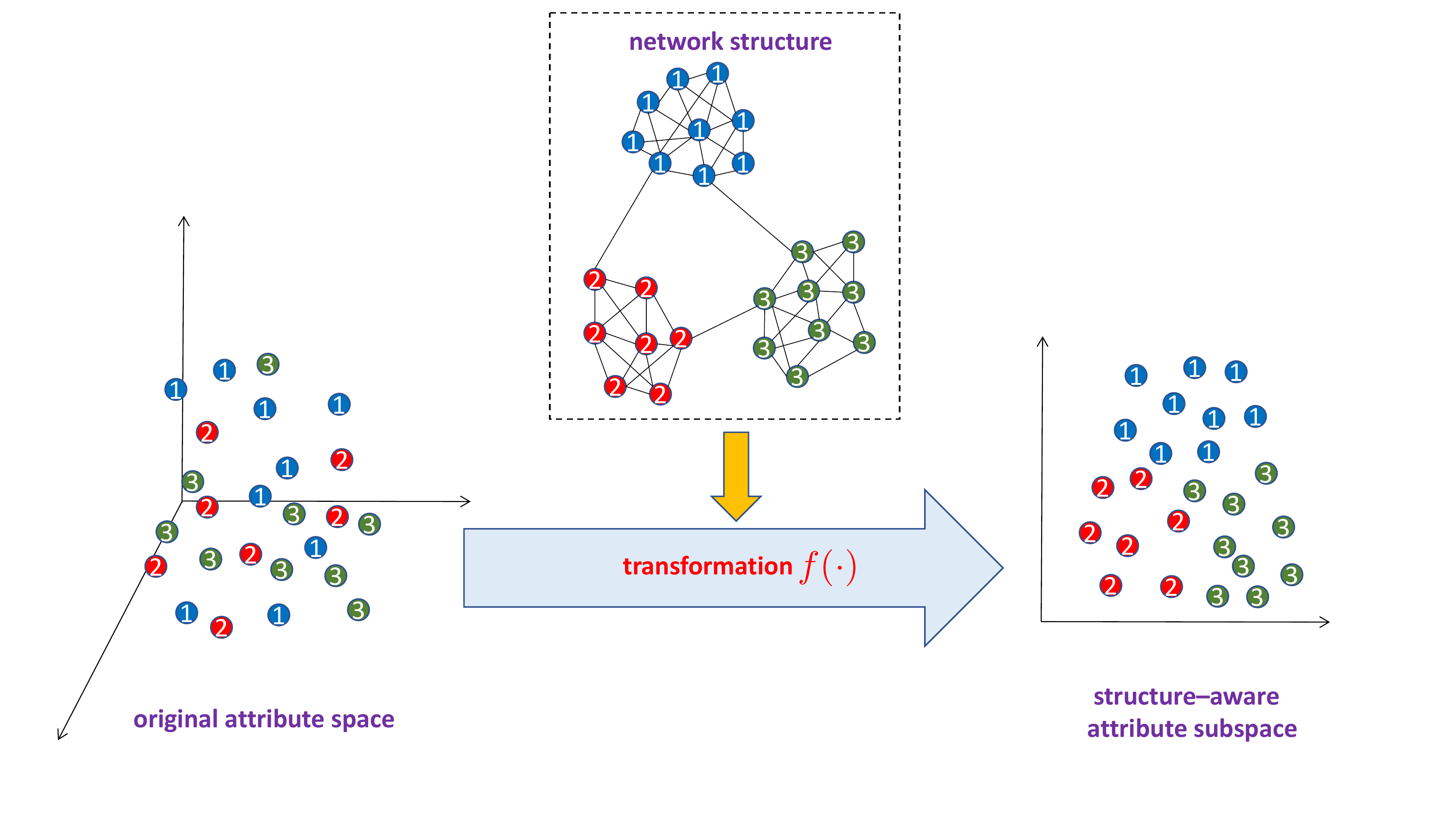}
	\caption{The working mechanism of the proposed attri2vec algorithm. A transformation $f(\cdot)$ guided by network structure is performed from the original node attribute space to seek a structure-aware attribute subspace, where node attributes and network structure can better compliment each other in a more consistent way towards learning high-quality node representations. }
	\label{fig:illustration} 
\end{figure*}

Motivated by aforementioned observations, we propose a unified attributed network embedding framework, called attri2vec, which learns node representations by discovering the \textit{latent subspace} of node attributes that well respects network structure. The working mechanism of attri2vec is illustrated in Fig.~\ref{fig:illustration}. The core idea is to perform a transformation $f(\cdot)$ guided by network structure from the original node attribute space, where node attribute distributions are not well aligned with network structure, to a structure-aware attribute subspace, where structurally similar nodes are located close to each other. The latent attribute subspace can better respect network structure in a consistent way towards learning high-quality node representations. Following~\citep{perozzi2014deepwalk}, attri2vec generates random walks to capture structural context. After random walks are generated, attri2vec uses the image $f(\bm{x}_{i})$ of node $v_{i}$ with attribute $\bm{x}_{i}$ in the new attribute subspace to predict its context nodes collected from random walk sequences. In this way, network structure is seamlessly encoded into the new attribute subspace by allowing nodes sharing similar neighbors to be located closely to each other. As the attribute subspace also preserves node attribute proximity, the learned node representations are informative enough to capture both the structure-level and node content-level proximity. As attri2vec infers attribute subspace that well respects network structure through directly learning the transformation $f(\cdot)$ from data, the weights for the respective contributions made by attributes and network structure for embedding learning is automatically learned from the data itself, which avoids the laborious weight parameter turning. In our prior work~\citep{zhang2017user}, we have proposed the UPP-SNE algorithm that learns user embeddings in social networks by leveraging both the network structure and user profile features. It constructs user embeddings by performing a non-linear kernel mapping~\citep{rahimi2008random} on user profile features, in which noisy information in user profiles is filtered out. Initial experimental results on social networks have shown that user embeddings learned by UPP-SNE can achieve substantial performance gains in node classification and clustering tasks. 

In this work,  we generalize the idea of UPP-SNE to tackle generic network embedding problems and also to cope with large-scale networks. First, we investigate four different types of linear or non-linear transformation functions for discovering the latent, structure-aware attribute subspace, and empirically validate their effectiveness on various types of networks. Second, we develop an efficient stochastic gradient descent algorithm to solve the optimization problem, which makes attri2vec able to be efficiently trained in an online mode. At each iteration, after sampling a node context pair, we only update the parameters related to the corresponding partial objective with gradient descent. Compared with the gradient descent strategy used by UPP-SNE, the stochastic gradient descent is not only more efficient, but also able to effectively mitigate the local minima problem, thus discovering more accurate solutions for attri2vec. Third, we further validate the ability of attri2vec in handling both binary and continuous attribute inputs for embedding learning. Fourth, given that attri2vec learns a network structure aware mapping on node attributes, it provides an alternative way to solve the out-of-sample problem. In other words, on the dynamically evolving networks, when new nodes join in, rather than learning representations for all nodes from the scratch, attri2vec can construct representations for new coming nodes from their available node attributes via the mapping function learned from previous network snapshots. We also conduct node classification and link prediction experiments to study the effectiveness of attri2vec in solving the out-of-sample problem. 


The contribution of this paper can be summarized as follows:
\begin{itemize}
	\item[(1)] We propose a unified framework for attributed social network embedding, attri2vec, that learns node embeddings by discovering a latent, structure aware attribute subspace and formulate an optimization problem.
	\item[(2)] We develop an efficient stochastic gradient descent algorithm to solve the optimization problem, which not only remedies the local minima problem, but also improves the effectiveness and efficiency, thus making attri2vec able to scale to large-scale networks.
	\item[(3)] We validate the effectiveness and efficiency of attri2vec on five real-world networks of various types. Extensive experiments on multi-class node classification, node clustering and link prediction tasks demonstrate that attri2vec learns node embeddings of better quality than state-of-the-art baselines, with an additional advantage to solve the out-of-sample problem.
\end{itemize}

The remainder of this paper is organized as follows. In Section~\ref{sec:related}, we review related works on network embedding. In Section~\ref{sec:probandpre}, we give a formal definition of the attributed network embedding problem and review the preliminaries of DeepWalk. The proposed attri2vec framework is then detailed in Section~\ref{sec:attri2vec}, followed by the experiments presented in Section~\ref{sec:experiments}. Finally, we conclude this paper in Section~\ref{sec:conclusion}.

\section{Related Work}
\label{sec:related}
Existing research work on network embedding can be roughly divided into two categories~\citep{zhang2018network}: structure preserving network embedding that leverages only network structure, and attributed network embedding that couples network structure with node attributes to improve network embedding.

\subsection{Structure Preserving Network Embedding} DeepWalk~\citep{perozzi2014deepwalk} is the pioneer work on network embedding, which generalizes the idea of word2vec~\citep{mikolov2013distributed} that learns word embeddings with the assumption that two words sharing similar context tend to have similar meaning. DeepWalk exploits Skip-Gram model~\citep{mikolov2013distributed} to characterize the node-context relations captured by random walks. To better balance local and global structure preserving, node2vec~\citep{grover2016node2vec} (a variant of DeepWalk) exploits biased random walk to generate node context. Different from random walk based methods, LINE \citep{tang2015line} learns node embeddings by directly modeling the first-order proximity (the proximity between connect nodes) and the second-order proximity (the proximity between nodes sharing common neighbors). GraRep~\citep{cao2015grarep} steps further to consider the high-order proximities by modeling the relations between nodes and their $k$-step neighbors with the matrix factorization version of Skip-Gram \citep{levy2014neural}. To learn deep, highly non-linear node representations, deep learning techniques are adopted by DNGR~\citep{cao2016deep} and SDNE~\citep{wang2016structural}. DNGR~\citep{cao2016deep} firstly calculates the positive pointwise mutual information (PPMI) matrix representation that well preserves the network structure, then feeds each row of the PPMI matrix \citep{levy2014neural} as input into the SDAE (Stacked Denoising Autoencoder) \citep{vincent2010stacked} to learn deep low-dimensional node embeddings. SDNE~\citep{wang2016structural} learns node embeddings via a semi-supervised autoencoder, which simultaneously preserves the first-order and the second-order proximity. M-NMF~\citep{wang2017community} complements the local structural proximity with community structure via modularity maximization~\citep{newman2006finding}. The structure preserving network embedding algorithms leverage only network structure to learn node embeddings and do not exert the power of widely available node content information.

\subsection{Attributed Network Embedding}

TADW~\citep{yang2015network} is the first attempt to incorporate rich node textual features into network embedding. TADW proves the equivalence between DeepWalk and a matrix factorization formulation, and then encodes node text features into the matrix factorization process for learning more informative node representations. To enforce TADW with the first-order proximity, HSCA \citep{zhang2016homophily} adds a graph regularization term on the objective of TADW to penalize the distance of connected nodes in the embedding space. Though TADW and HSCA are effective in leveraging rich node text features for learning informative node embeddings, their working mechanism is not so straightforward, i.e., they do not provide a clear objective to interpret how network structure and node attributes interplay with each other. Recently, some interpretable attributed network embedding algorithms have been proposed. To effectively leverage useful information in noisy, sparse and incomplete user profile features, UPP-SNE \citep{zhang2017user} learns node embeddings by performing a network structure aware non-linear mapping on user profile features. The SNE algorithm~\citep{liao2018attributed} learns node embeddings for attributed networks through a neural network, in which, for each node, its embedding is aggregated from its ID embedding that captures the structural proximity and attribute embedding that captures the attribute proximity. MVC-DNE~\citep{yang2017properties} applies deep autoencoder to node adjacent matrix representations and node content attributes respectively, and adopts cross-view learning to capture the interplay between network structure and node content. GraphSAGE~\citep{hamilton2017inductive} infers node representations through node content features by iteratively aggregating representations of neighboring nodes. AANE~\citep{huang2017accelerated} learns node representations by performing symetric matrix factorization~\citep{kuang2012symmetric} on attribute affinity matrix, and simultaneously minimizing the representation difference between connected nodes. However, the performance of existing attributed network embedding would be largely degraded in real-world scenarios where node attributes are sparse and incomplete, and in particular, when there exists a discrepancy between network structure and node attributes.
 
In addition to the above unsupervised network embedding algorithms, there are also some research works on supervised network embedding, like DDRW~\citep{li2016discriminative}, DMF~\citep{zhang2016collective}, TriDNR~\citep{pan2016tri}, LANE~\citep{huang2017label}. The supervised network embedding algorithms encode label information into the network embedding process. The learned embeddings not only well respect the network structure and node content attributes, but also possess the discriminative power for more accurate node classification.  Depending on the existence of node labels, the supervised network embedding is only tailored for node classification task and cannot be generalized to other tasks. In this work, we focus on the unsupervised attributed network embedding, which leverages network structure and node attributes to learn task-general node embeddings.

\section{Problem Definition and Preliminaries}
\label{sec:probandpre}
In this section, we give a formal problem definition of attributed network embedding and the preliminary background about DeepWalk.

\subsection{Problem Definition}
We assume that an attributed information network is given as $G=(V,E,X)$, with $V$ denoting the set of nodes and $E\subseteq V\times V$ denoting the set of edges. In $G$, $X\in\mathbb{R}^{m\times|V|}$ is the node attribute matrix, with the $i$-th column of $X$, $\bm{x}_{i}\in\mathbb{R}^{m}$, denoting the $m$-dimensional content feature vector for node $v_{i}\in V$.  By taking advantage of network structure and node attributes, the task of attributed network embedding is to effectively embed nodes into a low-dimensional vector space so as to obtain the node images $\mathrm{\Phi}(v_{i})\in\mathbb{R}^{d}$ in the latent space as node vector-format representations.

The learned vector-format node representations $\mathrm{\Phi}(v_{i})$ are expected to embody the following properties: (1) \textit{low-dimensional}, for the efficiency of the subsequent analytic tasks, the dimension of $\mathrm{\Phi}(v_{i})$, $d$, should be much smaller than the dimension of the original adjacency matrix representation, $|V|$; (2) \textit{structure preserving}, network structure is well encoded into node representations, i.e., two nodes with structural proximity should be embedded closely in the embedding space, (3) \textit{attribute preserving}, node attribute proximity should be well captured so that it complements rather than deteriorates network structure.

\subsection{DeepWalk}
For learning node embeddings, DeepWalk~\citep{perozzi2014deepwalk} exploits random walk to explore node neighborhood structure. By taking an analogy between truncated random walk sequences and sentences in natural language, DeepWalk extends the Skip-Gram model~\citep{mikolov2013distributed} from word representation learning to network embedding, with the expectation that two nodes sharing similar neighborhood structure have similar low-dimensional representations. Given a random walk node sequence $\mathcal{S}=\{v_{r_{1}},v_{r_{2}},\cdots,v_{r_{|\mathcal{S}|}}\}$ with length $|\mathcal{S}|$, DeepWalk learns embedding $\mathrm{\Phi}(v_{r_{i}})$ for $v_{r_{i}} (1\leq i\leq L)$, by using $\mathrm{\Phi}(v_{r_{i}})$ to predict $v_{r_{i}}$'s context in $t$-window size:
\begin{equation}
\max_{\mathrm{\Phi}}\log\mathrm{Pr}(\{v_{r_{i-t}},\cdots,v_{r_{i+t}}\}\setminus v_{i}|\mathrm{\Phi}(v_{r_{i}})).
\end{equation}
By exploiting the conditional independence assumption, the probability\\ ${\mathrm{Pr}(\{v_{r_{i-t}},\cdots,v_{r_{i+t}}\}\setminus v_{r_{i}}|\mathrm{\Phi}(v_{r_{i}}))}$ can be calculated as
\begin{equation}
\mathrm{Pr}(\{v_{r_{i-t}},\cdots,v_{r_{i+t}}\}\setminus v_{r_{i}}|\mathrm{\Phi}(v_{r_{i}}))=\prod_{j=i-t,j\neq i}^{i+t}\mathrm{Pr}(v_{r_{j}}|\mathrm{\Phi}(v_{r_{i}})).
\end{equation}
To model the probability $\mathrm{Pr}(v_{r_{j}}|\mathrm{\Phi}(v_{r_{i}}))$, an one-hidden-layer neural network is utilized with the outputted probability:
\begin{equation}\label{pair_prob2}
\mathrm{Pr}(v_{r_{j}}|\mathrm{\Phi}(v_{r_{i}}))=\frac{\exp(\mathrm{\Phi}(v_{r_{i}})\cdot\bm{w}^{out}_{r_{j}})}{\sum_{k=1}^{|V|}\exp(\mathrm{\Phi}(v_{r_{i}})\cdot\bm{w}^{out}_{k})},
\end{equation}where node embedding $\mathrm{\Phi}(v_{r_{i}})$ is the representation in the hidden layer, and $\bm{w}^{out}_{r_{j}}$ is the $r_{j}$-th column of $W^{out}\in\mathbb{R}^{d\times|V|}$ (the weight matrix from the hidden layer to the output layer). The node embedding $\mathrm{\Phi}(v_{i})$ is constructed as
\begin{equation}
\mathrm{\Phi}(v_{r_{i}})={W^{in}}^{\mathrm{T}}\bm{p}_{r_{i}}=\bm{w}^{in}_{r_{i}},
\end{equation}where $\bm{p}_{r_{i}}\in\mathbb{R}^{|V|}$ is the one-hot representation of $v_{r_{i}}$ in the input layer, with the value of $r_{i}$-th dimension being 1 and the values of other dimensions being 0, and $\bm{w}^{in}_{r_{i}}$ is the transpose of the $r_{i}$-th row of $W^{in}\in\mathbb{R}^{|V|\times d}$ (the weight matrix from the input layer to hidden layer).


\section{The attri2vec Framework}\label{sec:attri2vec}
In DeepWalk, the node embeddings $\mathrm{\Phi}(v_{i})$ are learned from scratch, which is independent of node attributes. However, as proved in previous work~\citep{yang2015network,zhang2016homophily}, node content features are crucially useful for enhancing the only structure preserving network embedding. To obtain more informative node embeddings, here, we adapt the DeepWalk's architecture to capture both network structure and node content features. 

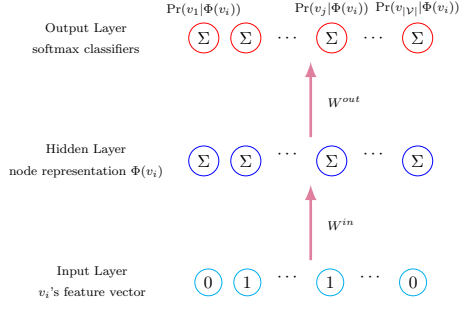
\begin{figure}
	\centering
	\begin{tikzpicture}
	[thick,scale=0.6, every node/.style={scale=0.6}]
	\matrix[nodes={red, draw, line width=0.2pt}, column sep=0.15cm](O1)
	{
		\node[circle] (O11) {\large\color{black}$\mathrm{\Sigma}$} ; &
		\node[circle] (O12) {\large\color{black}$\mathrm{\Sigma}$} ; &
		\node[draw=none,fill=none] {\large\color{black}$\cdots$}; &
		\node[circle] (O1j) {\large\color{black}$\mathrm{\Sigma}$} ; &
		\node[draw=none,fill=none] {\large\color{black}$\cdots$}; &
		\node[circle] (O1V) {\large\color{black}$\mathrm{\Sigma}$} ; \\
	};
	\matrix[below = 1cm of O1, nodes={blue, draw, line width=0.2pt}, column sep=0.15cm](H)
	{
		\node[circle] (H1) {\large\color{black}$\mathrm{\Sigma}$} ; &
		\node[circle] (H2) {\large\color{black}$\mathrm{\Sigma}$} ; &
		\node[draw=none,fill=none] {\large\color{black}$\cdots$}; &
		\node[circle] (Hj) {\large\color{black}$\mathrm{\Sigma}$} ; &
		\node[draw=none,fill=none] (Hdots) {\large\color{black}$\cdots$}; &
		\node[circle] (HV) {\large\color{black}$\mathrm{\Sigma}$} ; \\
	};
	\matrix[below = 1cm of H, nodes={cyan, draw, line width=0.2pt}, column sep=0.15cm](I)
	{
		\node[circle] (I1) {\large\color{black}0} ; &
		\node[circle] (I2) {\large\color{black}1} ; &
		\node[draw=none,fill=none] {\large\color{black}$\cdots$}; &
		\node[circle,fill=none] (Ij) {\large\color{black}1} ; &
		\node[draw=none,fill=none] (Idots) {\large\color{black}$\cdots$}; &
		\node[circle] (IV) {\large\color{black}0} ; \\
	};
	
	\draw [purple!50,line width=1pt,-latex] (I) -- (H);
	\draw [purple!50,line width=1pt,-latex] (H) -- (O1);
	
	\matrix[draw=none,fill=none, left = 0.1cm of O11, row sep=0.01cm](Out)
	{
		\node[text width=12em, text centered]{Output Layer};\\
		\node[text width=12em, text centered]{softmax classifiers};\\
	};
	
	\matrix[draw=none,fill=none, left = 0.1cm of H1, row sep=0.01cm](Hin)
	{
		\node[text width=12em, text centered]{Hidden Layer};\\
		\node[text width=12em, text centered]{node representation $\mathrm{\Phi}(v_{i})$};\\
	};
	
	\matrix[draw=none,fill=none, left = 0.1cm of I1, row sep=0.01cm](In)
	{
		\node[text width=12em, text centered]{Input Layer};\\
		\node[text width=12em, text centered]{$v_{i}$'s feature vector};\\
	};
	
	\node[draw=none, fill=none, above = 0.02cm of O11, text width=8em, text centered] {\small $\mathrm{Pr}(v_{1}|\mathrm{\Phi}(v_{i}))$};
	\node[draw=none, fill=none, above = 0.02cm of O1j, text width=8em, text centered]  {\small $\mathrm{Pr}(v_{j}|\mathrm{\Phi}(v_{i}))$};
	\node[draw=none, fill=none, above = 0.02cm of O1V, text width=8em, text centered] {\small $\mathrm{Pr}(v_{|\mathcal{V}|}|\mathrm{\Phi}(v_{i}))$};
	
	\node[draw=none, fill=none, above left = 0.45cm and -0.6cm of Idots, text width=8em, text centered] {\small $W^{in}$};
	\node[draw=none, fill=none, above left = 0.45cm and -0.6cm of Hdots, text width=8em, text centered] {\small $W^{out}$};
	
	\end{tikzpicture}
	\caption{The architecture of attri2vec. For each node context pair $(v_{i},v_{j})$, attri2vec learns node representations by modeling $\mathrm{Pr}(v_{j}|\mathrm{\Phi}(v_{i}))$. attri2vec firstly constructs node representations in the hidden layer by performing a linear or non-linear transformation on $v_{i}$'s content attributes, then uses the hidden layer representation to predict the probability $\mathrm{Pr}(v_{j}|\mathrm{\Phi}(v_{i}))$ with softmax.}
	\label{fig:mechanism} 
\end{figure}

To learn node embeddings $\mathrm{\Phi}(\cdot)$, DeepWalk solves the following joint optimization problem:
\begin{equation}\label{optimization_prob}
\min_{{W^{in},W^{out}}}\mathcal{O}(W^{in},W^{out}),
\end{equation}where
\begin{equation}
\begin{aligned}
&\mathcal{O}(W^{in},W^{out})\\
&=-\sum_{\mathcal{S}}\sum_{i=1}^{|\mathcal{S}|}\log\mathrm{Pr}(\{v_{r_{i-t}},\cdots,v_{r_{i+t}}\}\setminus v_{r_{i}}|\mathrm{\Phi}(v_{i})),\\
&=-\sum_{\mathcal{S}}\sum_{i=1}^{|\mathcal{S}|}\sum_{j=i-t,j\neq i}^{i+t}\mathrm{Pr}(v_{r_{j}}|\mathrm{\Phi}(v_{r_{i}})).
\end{aligned}
\end{equation}
For convenience, we rewrite the objective function $\mathcal{O}(W^{in},W^{out})$ as
\begin{equation}\label{obj}
\mathcal{O}(W^{in},W^{out})=-\sum_{i=1}^{|V|}\sum_{j=1}^{|V|}n(v_{i},v_{j})\log\mathrm{Pr}(v_{j}|\mathrm{\Phi}(v_{i})),
\end{equation}where $n(v_{i},v_{j})$ is the number of times that $v_{j}$ occurs in $v_{i}$ context within $t$-window size in the generated set of random walks. Using Eq. (\ref{pair_prob2}), we model the probability $\mathrm{Pr}(v_{j}|\mathrm{\Phi}(v_{i}))$ as
\begin{equation}\label{pair_prob3}
\mathrm{Pr}(v_{j}|\mathrm{\Phi}(v_{i}))=\frac{\exp(\mathrm{\Phi}(v_{i})\cdot\bm{w}^{out}_{j})}{\sum_{k=1}^{|V|}\exp(\mathrm{\Phi}(v_{i})\cdot\bm{w}^{out}_{k})}.
\end{equation}Here, different from DeepWalk that constructs $\mathrm{\Phi}(v_{i})$ with a linear transformation from node $v_{i}$'s one-hot representation $\bm{p}_{i}$, to incorporate node content features, we construct $\mathrm{\Phi}(v_{i})$ from node $v_{i}$'s content features $\bm{x}_{i}\in\mathbb{R}^{m}$ with a linear or non-linear transformation:
\begin{equation}
\mathrm{\Phi}(v_{i})=f(\bm{x}_{i}),
\end{equation}To obtain the embedding $\mathrm{\Phi}(v_{i})$ for each node $v_{i}\in V$, a series of mapping function $f(\cdot):\mathbb{R}^{m}\rightarrow\mathbb{R}^{d}$ are investigated:
\begin{itemize}
	\item [(1)] \textbf{linear mapping}:
	\begin{equation}
	\begin{aligned}
	f(\bm{x}_{i})&={W^{in}}^{\mathrm{T}}\bm{x}_{i}\\
	&=\left[{\bm{w}^{in}_{1}}^{\mathrm{T}}\bm{x}_{i},\cdots,{\bm{w}^{in}_{d}}^{\mathrm{T}}\bm{x}_{i}\right]^{\mathrm{T}};
	\end{aligned}
	\end{equation}
	\item [(2)] \textbf{rectified linear unit (ReLU) mapping}:
	\begin{equation}
	f(\bm{x}_{i})=\left[\max(0,{\bm{w}^{in}_{1}}^{\mathrm{T}}\bm{x}_{i}),\cdots,\max(0,{\bm{w}^{in}_{d}}^{\mathrm{T}}\bm{x}_{i})\right]^{\mathrm{T}};
	\end{equation}
	\item [(3)] \textbf{approximated kernel mapping} used in UPP-SNE~\citep{zhang2017user}:
	\begin{equation}
	\begin{aligned}
	f(\bm{x}_{i})=\frac{1}{\sqrt{m}}&\left[\cos({\bm{w}^{in}_{1}}^{\mathrm{T}}\bm{x}_{i}),\cdots,\cos({\bm{w}^{in}_{d/2}}^{\mathrm{T}}\bm{x}_{i}),\right.\\
	&\left.\sin({\bm{w}^{in}_{1}}^{\mathrm{T}}\bm{x}_{i}),\cdots,\sin({\bm{w}^{in}_{d/2}}^{\mathrm{T}}\bm{x}_{i})\right]^{\mathrm{T}};
	\end{aligned}
	\end{equation}
	\item [(4)] \textbf{sigmoid mapping}:
	\begin{equation}
	\begin{aligned}
	f(\bm{x}_{i})=&\left[1/(1+\exp(-{\bm{w}^{in}_{1}}^{\mathrm{T}}\bm{x}_{i})),\cdots,\right.\\
	&\left.1/(1+\exp(-{\bm{w}^{in}_{d}}^{\mathrm{T}}\bm{x}_{i}))\right]^{\mathrm{T}}.
	\end{aligned}
	\end{equation}
\end{itemize}Above, $W^{in}\in\mathbb{R}^{m\times d}$ is the input-hidden weight matrix and $\bm{w}^{in}_{j}$ is the $j$-th column of $W^{in}$. The architecture of attri2vec is given in Fig \ref{fig:mechanism}. 
We denote the attri2vec algorithm with the four mapping functions as attri2vec-linear, attri2vec-ReLU, attri2vec-kernel and attri2vec-sigmoid respectively.

We adopt the stochastic gradient descent to solve the optimization problem in Eq. (\ref{obj}). After randomly sampling a node context pair $(v_{i},v_{j})$ according to the frequency distribution in $n(v_{i},v_{j})$, we try to update parameters for reducing the value the following partial objective:
\begin{equation}\label{partial_obj1}
\begin{aligned}
&\mathcal{O}_{ij}(W^{in},W^{out})=-\log\mathrm{Pr}(v_{j}|\mathrm{\Phi}(v_{i}))\\
&=-\mathrm{\Phi}(v_{i})\cdot\bm{w}^{out}_{j}+\log\sum_{k=1}^{|V|}\exp(\mathrm{\Phi}(v_{i})\cdot\bm{w}^{out}_{k}).
\end{aligned}
\end{equation}In Eq. (\ref{partial_obj1}), the calculation for the partial objective $\mathcal{O}_{ij}$ involves the summation over all nodes in the given network, which is prohibitively time-consuming. To make a speed-up, we adopt negative sampling~\citep{gutmann2012noise} to approximate the partial objective:
\begin{equation}\label{partial_obj2}
\begin{aligned}
\mathcal{O}_{ij}(W^{in},W^{out})&=-\log\sigma(\mathrm{\Phi}(v_{i})\cdot\bm{w}^{out}_{j})\\
&-\sum_{k=1}^{K}\log\sigma(-\mathrm{\Phi}(v_{i})\cdot\bm{w}^{out}_{N_{i,k}}),
\end{aligned}
\end{equation}where $\sigma(\cdot)$ is the sigmoid function with $\sigma(x)=1/(1+\exp(-x))$, $N_{i,k}$ is the index for the $k$-th negative node sampled for node $v_{i}$ and $K$ is the number of sampled negative nodes. Exploiting the approximated partial objective $\mathcal{O}_{ij}(W^{in},W^{out})$ in Eq. (\ref{partial_obj2}), the parameters are updated by:
\begin{equation}\label{update}
\begin{aligned}
&\bm{w}^{in}_{p} = \bm{w}^{in}_{p} - \eta\frac{\partial\mathcal{O}_{ij}}{\partial\bm{w}^{in}_{p}}, \quad p\in\{1,2,\cdots,d\};\\
&\bm{w}^{out}_{q} = \bm{w}^{out}_{q} - \eta\frac{\partial\mathcal{O}_{ij}}{\partial\bm{w}^{out}_{q}},\quad q\in\{j\}\cup\{N_{i,1},N_{i,2},\cdots,N_{i,K}\}.
\end{aligned}
\end{equation}

The gradients are calculated as
\begin{equation}
\begin{aligned}
\frac{\partial \mathcal{O}_{ij}}{\bm{w}^{in}_{p}}&=-\sigma(-\mathrm{\Phi}(v_{i})\cdot\bm{w}^{out}_{j})\frac{\partial \mathrm{\Phi}(v_{i})}{\partial\bm{w}^{in}_{p}}\bm{w}^{out}_{j}\\
&+\sum_{k=1}^{K}\sigma(\mathrm{\Phi}(v_{i})\cdot\bm{w}^{out}_{N_{i,k}})\frac{\partial \mathrm{\Phi}(v_{i})}{\partial\bm{w}^{in}_{p}}\bm{w}^{out}_{N_{i,k}};
\end{aligned}
\end{equation}
\begin{equation}
\begin{aligned}
\frac{\partial \mathcal{O}_{ij}}{\bm{w}^{out}_{q}}&=-\mathbf{1}_{q}(j)\sigma(-\mathrm{\Phi}(v_{i})\cdot\bm{w}^{out}_{j})\mathrm{\Phi}(v_{i})\\
&+\sum_{k=1}^{K}\mathbf{1}_{q}(N_{i,k})\sigma(\mathrm{\Phi}(v_{i})\cdot\bm{w}^{out}_{N_{i,k}})\mathrm{\Phi}(v_{i}).
\end{aligned}
\end{equation}where $\frac{\partial\mathrm{\Phi}(v_{i})}{\partial\bm{w}^{in}_{p}}$ is the $m\times d$ Jacobian matrix under different mappings for constructing $\mathrm{\Phi}(v_{i})$. $\mathbf{1}_{q}(\cdot)$ is an indicator function, which is defined as
\begin{equation}
\mathbf{1}_{q}(x)=\left\lbrace
\begin{aligned}
&1\;\;\mathrm{if}\;x=q,\\
&0\;\;\mathrm{if}\;x\neq q.
\end{aligned}
\right.
\end{equation}

\begin{algorithm}[t]
	\caption{attri2vec: a Unified Framework for Attributed Network Embedding}
	\label{alg:attri2vec}
	\begin{algorithmic}[1]
		\REQUIRE ~~\\
		An attributed network $G=(V,E,X)$;
		\ENSURE ~~\\
		Node embedding $\mathrm{\Phi}(\cdot)$ for each $v_{i}\in V$;
		\STATE $\mathbb{S}$ $\leftarrow$ generate a set of random walks on $G$; \label{code:randwalk}
		\STATE $n(v_i,v_j)$ $\leftarrow$ count frequency of node context pairs ($v_{i},v_{j})$ in $\mathbb{S}$; \label{code:frequency}
		\STATE{$\{W^{in},W^{out}\} \leftarrow$ initialize parameters;} \label{code:initialization}
		\REPEAT \label{code:repeat}
		\STATE{$(v_i,v_j) \leftarrow$ sample a node context pair according to the distribution of $n(v_i,v_j)$;} \label{code:sample}
		\STATE{$\{v_{N_{i,1}},\cdots,v_{N_{i,K}}\}\leftarrow$ draw $K$ negative nodes;} \label{code:netagtive}
		\STATE{$\{W^{in},W^{out}\} \leftarrow$ update parameters with Eq.~(\ref{update});} \label{code:update}
		\UNTIL {maximum number of iterations expire;}\label{code:iteration}
		\STATE construct node embedding $\mathrm{\Phi}(\cdot)$ with $W^{in}$ and the selected mapping function; \label{code:embed}
		\STATE \textbf{return} $\mathrm{\Phi}(\cdot)$;
	\end{algorithmic}
\end{algorithm} 

The workflow of the attri2vec algorithm is given in Algorithm~\ref{alg:attri2vec}. Firstly, in line \ref{code:randwalk}, a set of random walk sequences are generated on the given attributed network $G$, by starting random walk with length $l$ from each node for $\gamma$ times. Then, in line \ref{code:frequency}, the statistics $n(v_{i},v_{j})$ is calculated from the generated random walks with a fixed window size $t$. In line \ref{code:initialization}, we initialize $W^{in}$ with random numbers, and initialize $W^{out}$ with 0. After that, in line \ref{code:repeat}-\ref{code:iteration}, the parameters are updated iteratively with stochastic gradient descent. At each iteration, a node context pair $(v_{i},v_{j})$ is sampled from the distribution of $n(v_{i},v_{j})$ with the alias table method \citep{li2014reducing}, which takes only $O(1)$ time; parameters are then updated for reducing the value of the partial objective $O_{ij}$ in Eq. (\ref{partial_obj2}). 

By leveraging the sparsity of node attributes, the time complexity of line \ref{code:update} is only $O(\bar{m}d)$ with $\bar{m}$ being the averaged number of non-zero elements of $\bm{x}_{i}$. The number of iterations is at the scale of the number of non-zero $n(v_{i},v_{j})$, which is up bounded by $2\gamma lt|V|$.  Taking $\gamma$, $l$ and $t$ as constants, attri2vec has an overall time complexity of $O(\bar{m}d|V|)$, which is linear to the number of nodes. This guarantees its efficiency and scalability over large-scale networks. 
\section{Experiments}
\label{sec:experiments}
In this section, we report experimental results on various types of real-world networks to evaluate the effectiveness and efficiency of the proposed attri2vec algorithm.

\subsection{Benchmark Networks}
\label{sec:experiments:networks}
In our experiments, five real-world attributed networks are used. Their details are as following:

\begin{itemize}
	\item \textbf{Citeseer}. The Citeseer network\footnote{\url{https://linqs.soe.ucsc.edu/data}\label{fn:snapdata}} includes 3,312 scientific publications from six categories. There exists 4,732 citations among these papers. Each paper is represented by a 3,703-dimensional binary vector, with each entry representing the presence/absence of the corresponding word. 
	\item \textbf{DBLP}. The DBLP network is a subgraph of the DBLP bibliographic network\footnote{\url{https://aminer.org/citation} (Version 3 is used)}. To construct the DBLP network, we extract papers from the four research areas: \textit{Database}, \textit{Data Mining}, \textit{Artificial Intelligence}, \textit{Computer Vision}, according to papers' venue information and remove papers with no citations. The DBLP network contains 18,448 papers and 45,661 citations. From paper titles, we construct 2,476-dimensional binary node feature vectors, with each element indicating the presence/absence of the corresponding word. 
	\item \textbf{PubMed}. The PubMed network\footref{fn:snapdata} is composed of 19,717 scientific diabetes publications in three categories: Diabetes Mellitus Experimental, Diabetes Mellitus Type 1, Diabetes Mellitus Type 2, together with 44,338 citation relations. Each publication is described by a TF-IDF weighted word vector from a dictionary formed by 500 unique words.
	\item \textbf{Facebook}. The Facebook network is merged from 10 Facebook ego-networks from SNAP\footnote{\url{https://snap.stanford.edu/data/}}. There are 4,039 users and 88,234 friendship relations. Each user is described by a 1403-dimensional bag-of-words vector from tree-structured user profiles~\citep{leskovec2012learning}. Users' education types are used as labels.
	\item \textbf{Flickr}. The Flickr network\footnote{\url{http://people.tamu.edu/~xhuang/Code.html}} is extracted from the Flick online photo sharing platform, which includes 7,575 users and 239,738 follower-followee relations. These users join in nine predefined interest groups. Users' features are described by the tags of their images. Each user is represented by a 12,047-dimensional binary vector, according to the occurrence/absence of the corresponding tag.
\end{itemize}
For the above networks, the direction of links is ignored. Their statistics are summarized in Table \ref{dataset}.

\renewcommand\arraystretch{1.35}
\begin{table}[t]
	\begin{center}
		\tabcolsep 3pt
		\caption{Summary of Five Real-world Networks}
		\begin{tabular}{cccccc}
			\hline
			& Citeseer & DBLP & PubMed & Facebook & Flickr \\\hline
			$|V|$ & 3,312 & 18,448 & 19,717& 4,039 & 7,575 \\
			$|E|$  & 4,732 & 45,611& 44,338 & 88,234 & 239,738 \\
			$m$ & 3,703 & 2,476 & 500 & 1,403 & 12,047 \\
			$nnz(X)$ & 105,165 & 103,130 & 988,031 & 31,656 & 182,517 \\
			\# of Class & 6 & 4 & 3 & 4 & 9 \\\hline
		\end{tabular}
		\label{dataset}
	\end{center}
\end{table}

\subsection{Baseline Methods}

We compare the attri2vec algorithm (attri2vec-linear, attri2vec-ReLU, attri2vec-kernel and attri2vec-sigmoid) with the following baseline methods:
\begin{itemize}
	\item \textbf{DeepWalk}~\citep{perozzi2014deepwalk}/\textbf{node2vec}~\citep{grover2016node2vec}. They both learn node representations by preserving the similarity between nodes sharing similar contexts in random walks. node2vec is equivalent to DeepWalk with the default parameter setting $p=q=1$.
	\item \textbf{LINE-1}~\citep{tang2015line}. LINE-1 denotes the version of LINE that learns node representations by modeling the first-order proximity.
	\item \textbf{LINE-2}~\citep{tang2015line}. LINE-2 denotes the version of LINE that learns node representations by preserving the second-order proximity.
	\item \textbf{SDNE}~\citep{wang2016structural}. SDNE learns deep node representations with a semi-supervised autoencoder, which captures both the first-order and the second-order proximity.
	\item \textbf{TADW}~\citep{yang2015network}. TADW imports node text features into network embedding though inductive matrix factorization~\citep{natarajan2014inductive}.
	\item \textbf{SNE}~\citep{liao2018attributed}. SNE constructs node representations by aggregating structure preserving ID embedding and attribute embedding.
	\item \textbf{MVC-DNE}~\citep{yang2017properties}. MVC-DNE fuses network structure and node content into node embeddings through deep autoencoder  cross-view learning.
\end{itemize}
Among the above baseline methods, DeepWalk, LINE-1, LINE-2 and SDNE leverage only network structure to learn node representations, while TADW, SNE and MVC-DNE integrate network structure with node content features.
\subsection{Experimental Settings}

For DeepWalk and the proposed attri2vec algorithm, we set the length of random walks $l$ as 100, the number of walks starting at per node $\gamma$ as 40, and the window size $t$ as 10. 

For fair comparisons, DeepWalk is trained in a same way as attri2vec, where a set of node context pairs are first collected from random walks, and parameters are then updated with stochastic gradient descent, by sampling a node context pair at each iteration. Negative sampling is adopted by DeepWalk, LINE-1, LINE-2, and attri2vec, where the number of negative samples $K$ is set to 5 uniformly. For the four stochastic gradient descent based algorithms, we set the maximum number of iterations as 100 million, and gradually decrease the learning rate $\eta$ from $0.025$ to $2.5\times 10^{-6}$. When we run attri2vec-linear and attri2vec-ReLU on Flickr, to avoid gradient explosion, we set the initial learning rate to $0.005$, and gradually decrease the learning rate to $5\times 10^{-7}$. Parameters of TADW are set to their default values. For SDNE, its hyper parameter $\alpha$ and $\nu$ are both set to 0.01, and $\beta$ is set to 10. For SDNE, we set the number of neurons at each layer as 3312-128, and 18,448-512-128, 4039-128, 7575-128 for Citeseer, DBLP, Facebook and Flickr respectively. For MVC-DNE, on Citeseer, DBLP, Facebook and Flickr, in structure view, we respectively set the number of neurons at each layer as 3312-64, 18,448-256-64, 4039-64, and 7575-64; in the node attribute view, the number of neurons at each layer is respectively set to 3,703-64, 2,476-64, 1,403-64, and 12,047-256-64. For SDNE and MVC-DNE, 500 epochs are respectively run for pre-training and parameter fine-tuning. Other parameters of SDNE and MVC-DNE are set according to~\citep{yang2017properties}. For SNE, default parameters are used. For attri2vec and all baseline methods, the dimension of learned node representations is set to 128.

\subsection{Node Classification Experiments}

\begin{table*}[t]
	\centering
		\caption{Node Classification Results on Citeseer}
		\tabcolsep 3pt
		\renewcommand{\arraystretch}{1.1}
		\scalebox{0.85}{
		\begin{tabular}{clcccccccccc}
			\hline
			&Training Ratio & 10\% & 20\% & 30\% & 40\% & 50\% & 60\% & 70\% & 80\% & 90\% \\ \hline
			\multirow{11}{*}{Micro-$F_{1}$(\%)}& DeepWalk & 51.05 & 54.72 & 57.89 & 58.52 & 59.48 & 59.59 & 59.49 & 59.71 & 61.48 \\
			& LINE-1 & 43.73 & 48.51 & 50.17 & 52.10 & 53.22 & 54.03 & 54.67 & 54.24 & 56.77 \\
			& LINE-2 & 30.64 & 35.80 & 39.28 & 41.66 & 42.57 & 43.72 & 44.78 & 43.85 & 46.80 \\
			& SDNE & 34.96 & 38.97 & 42.20 & 44.13 & 44.46 & 45.48 & 45.69 & 44.29 & 45.95 \\
			& TADW & \underline{61.21} & \underline{64.05} & 64.88 & 65.20 & 65.72 & 65.65 & 65.84 & 65.94 & 67.22 \\
			& SNE & 32.31 & 37.61 & 43.46 & 46.33 & 47.23 & 48.81 & 48.68 & 49.91 & 50.73 \\
			& MVC-DNE & 50.73 & 54.95 & 59.81 & 63.09 & 64.57 & 65.92 & 66.90 & 67.07 & 68.07 \\
			& attri2vec-linear & 52.69 & 54.47 & 62.74 & 65.72 & 66.97 & 67.28 & 67.62 & 68.35 & 69.94 \\
			& attri2vec-ReLU & 53.47 & 55.82 & 61.94 & 65.71 & 67.25 & 67.69 & 68.04 & 68.56 & 71.06 \\
			& attri2vec-kernel & \textbf{62.04} & \textbf{65.87} & \textbf{67.56} & \underline{69.03} & \underline{70.21} & \underline{69.63} & \underline{70.16} & \underline{70.74} & \underline{72.24} \\
			& attri2vec-sigmoid & 59.47 & 60.51 & \underline{65.51} & \textbf{69.34} & \textbf{70.27} & \textbf{70.72} & \textbf{71.63} & \textbf{71.53} & \textbf{73.17} \\
			\hline
			\multirow{11}{*}{Macro-$F_{1}$(\%)} & DeepWalk & 47.74 & 51.23 & 53.52 & 53.73 & 54.08 & 53.87 & 53.44 & 53.54 & 55.41 \\
			& LINE-1 & 41.29 & 45.27 & 46.55 & 47.77 & 48.85 & 49.63 & 49.94 & 49.38 & 51.30 \\
			& LINE-2 & 28.64 & 33.29 & 36.06 & 37.73 & 38.20 & 38.88 & 39.83 & 38.78 & 41.78 \\
			& SDNE & 32.38 & 36.23 & 38.82 & 39.88 & 40.23 & 40.84 & 41.17 & 39.26 & 40.36 \\
			& TADW & 55.41 & \underline{57.95} & 58.75 & 59.14 & 59.60 & 60.16 & 60.04 & 59.95 & 60.76 \\
			& SNE & 30.30 & 35.40 & 39.69 & 42.11 & 42.30 & 44.06 & 43.51 & 44.39 & 44.81 \\
			& MVC-DNE & 47.36 & 51.18 & 55.26 & 57.76 & 58.52 & 60.10 & 60.20 & 60.79 & 60.97 \\
			& attri2vec-linear & 49.15 & 51.09 & 58.28 & 61.11 & 62.09 & 62.35 & 62.59 & 62.78 & 64.92 \\
			& attri2vec-ReLU & 49.92 & 52.48 & 57.79 & 61.53 & 62.40 & 63.04 & 63.08 & 63.55 & 65.58 \\
			& attri2vec-kernel & \textbf{58.12} & \textbf{61.81} & \textbf{63.36} & \underline{64.96} & \textbf{66.00} & \underline{65.50} & \underline{65.80} & \underline{66.27} & \textbf{68.39} \\
			& attri2vec-sigmoid & \underline{55.64} & 57.11 & \underline{61.41} & \textbf{65.05} & \underline{65.69} & \textbf{66.24} & \textbf{67.04} & \textbf{66.68} & \underline{68.24} \\
			\hline
		\end{tabular}}
		\label{Res_Classification_Citeseer}
\end{table*}

\begin{table*}[t]
	\centering
		\caption{Node Classification Results on DBLP}
		\tabcolsep 3pt
		\renewcommand{\arraystretch}{1.1}
		\scalebox{0.85}{
			\begin{tabular}{clcccccccccc}
				\hline
				&Training Ratio & 10\% & 20\% & 30\% & 40\% & 50\% & 60\% & 70\% & 80\% & 90\% \\ \hline
				\multirow{11}{*}{Micro-$F_{1}$(\%)}& DeepWalk & 78.58 & 79.92 & 80.08 & 80.30 & 80.42 & 80.34 & 80.46 & 80.44 & 80.30 \\
				& LINE-1 & 75.39 & 76.63 & 77.07 & 77.24 & 77.43 & 77.39 & 77.49 & 77.67 & 77.57 \\
				& LINE-2 & 68.08 & 69.76 & 70.23 & 70.47 & 70.81 & 70.53 & 70.89 & 70.65 & 70.81 \\
				& SDNE & 63.59 & 64.96 & 65.23 & 65.46 & 65.60 & 65.51 & 65.43 & 65.98 & 64.78 \\
				& TADW & 75.18 & 76.03 & 76.79 & 77.23 & 77.22 & 77.31 & 77.37 & 77.72 & 77.77 \\
				& SNE & 68.37 & 70.19 & 70.73 & 70.81 & 71.29 & 71.08 & 71.20 & 71.10 & 71.01 \\
				& MVC-DNE & 73.37 & 75.17 & 75.74 & 75.94 & 76.18 & 76.17 & 76.35 & 76.12 & 76.23 \\
				& attri2vec-linear & 76.42 & 78.36 & 78.68 & 78.97 & 79.21 & 79.05 & 79.32 & 79.11 & 79.23 \\
				& attri2vec-ReLU & \textbf{80.86} & \textbf{82.91} & \textbf{83.31} & \textbf{83.55} & \textbf{83.80} & \textbf{83.57} & \textbf{84.00} & \textbf{83.95} & \textbf{83.95} \\
				& attri2vec-lernel & 79.56 & 80.58 & 81.11 & 81.45 & 81.54 & 81.30 & 81.63 & 81.57 & 81.74 \\
				& attri2vec-sigmoid & \underline{79.89} & \underline{81.66} & \underline{82.16} & \underline{82.32} & \underline{82.52} & \underline{82.47} & \underline{82.72} & \underline{82.47} & \underline{82.78} \\
				\hline
				\multirow{11}{*}{Macro-$F_{1}$(\%)} & DeepWalk & 71.12 & 72.88 & 72.96 & 73.07 & 73.26 & 73.20 & 73.18 & 73.19 & 73.63 \\
				& LINE-1 & 66.91 & 68.41 & 68.95 & 69.18 & 69.36 & 69.38 & 69.41 & 69.59 & 70.03 \\
				& LINE-2 & 59.95 & 61.63 & 62.01 & 62.39 & 62.63 & 62.27 & 62.60 & 62.24 & 62.91 \\
				& SDNE & 49.27 & 49.90 & 49.52 & 49.74 & 50.07 & 49.39 & 49.19 & 49.99 & 48.91 \\
				& TADW & 65.17 & 67.14 & 68.03 & 68.47 & 68.61 & 68.65 & 68.50 & 68.91 & 69.61 \\
				& SNE & 57.78 & 59.80 & 60.17 & 59.87 & 60.73 & 60.37 & 60.40 & 60.09 & 60.93 \\
				& MVC-DNE & 65.04 & 67.26 & 67.76 & 67.68 & 67.96 & 68.00 & 68.23 & 67.97 & 67.98 \\
				& attri2vec-linear & 69.19 & 71.79 & 71.99 & 72.16 & 72.59 & 72.51 & 72.66 & 72.15 & 73.09 \\
				& attri2vec-ReLU & \textbf{73.99} & \textbf{77.03} & \textbf{77.27} & \textbf{77.62} & \textbf{77.85} & \textbf{77.71} & \textbf{77.94} & \textbf{78.07} & \textbf{78.61} \\
				& attri2vec-kernel & 73.04 & 74.51 & 75.13 & 75.51 & 75.62 & 75.38 & 75.60 & 75.65 & 76.28 \\
				& attri2vec-sigmoid & \underline{73.33} & \underline{75.97} & \underline{76.51} & \underline{76.64} & \underline{76.88} & \underline{77.05} & \underline{77.10} & \underline{76.89} & \underline{77.67} \\
				\hline
		\end{tabular}}
		\label{Res_Classification_DBLP}
\end{table*}

\begin{table*}[t]
	\centering
	\caption{Node Classification Results on PubMed}
	\tabcolsep 3pt
	\renewcommand{\arraystretch}{1.1}
	\scalebox{0.85}{
		\begin{tabular}{clcccccccccc}
			\hline
			&Training Ratio & 10\% & 20\% & 30\% & 40\% & 50\% & 60\% & 70\% & 80\% & 90\% \\ \hline
			\multirow{11}{*}{Micro-$F_{1}$(\%)}& DeepWalk & 39.58 & 39.69 & 39.59 & 39.49 & 39.67 & 39.85 & 39.37 & 39.92 & 40.65 \\
			& LINE-1 & 76.49 & 77.29 & 77.67 & 77.81 & 78.12 & 77.86 & 78.06 & 77.96 & 77.86 \\
			& LINE-2 & 75.19 & 76.19 & 76.60 & 76.78 & 77.19 & 76.77 & 76.97 & 76.98 & 77.21 \\
			& SDNE & 70.07 & 71.22 & 71.62 & 72.03 & 72.37 & 72.20 & 72.48 & 72.44 & 72.60 \\
			& TADW & 82.76 & 83.86 & 84.25 & 84.56 & 84.86 & 84.52 & 84.98 & 84.73 & 84.81 \\
			& SNE & 79.99 & 81.31 & 81.64 & 81.79 & 82.01 & 81.96 & 82.16 & 82.22 & 82.37 \\
			& MVC-DNE & 81.58 & 82.25 & 82.51 & 82.68 & 82.91 & 82.72 & 82.68 & 82.60 & 82.72 \\
			& attri2vec-linear & 84.39 & 85.66 & 86.06 & 86.10 & 86.46 & 86.30 & 86.69 & 86.54 & 86.31 \\
			& attri2vec-ReLU & \textbf{86.00} & \textbf{87.03} & \textbf{87.37} & \textbf{87.59} & \textbf{87.66} & \textbf{87.76} & \textbf{87.81} & \textbf{87.52} & \textbf{87.78} \\
			& attri2vec-kernel & 84.56 & 85.25 & 85.53 & 85.54 & 85.75 & 85.62 & 85.90 & 85.66 & 85.76 \\
			& attri2vec-sigmoid & \underline{85.96} & \underline{86.57} & \underline{86.80} & \underline{86.97} & \underline{86.97} & \underline{86.91} & \underline{87.17} & \underline{86.95} & \underline{86.99} \\
			\hline
			\multirow{11}{*}{Macro-$F_{1}$(\%)} & DeepWalk & 18.90 & 18.94 & 19.76 & 19.79 & 18.94 & 19.08 & 19.73 & 19.02 & 19.27 \\
			& LINE-1 & 75.03 & 75.89 & 76.23 & 76.40 & 76.74 & 76.47 & 76.71 & 76.59 & 76.37 \\
			& LINE-2 & 73.53 & 74.57 & 74.93 & 75.21 & 75.67 & 75.24 & 75.50 & 75.46 & 75.67 \\
			& SDNE & 67.50 & 68.50 & 68.85 & 69.35 & 69.75 & 69.52 & 69.68 & 69.72 & 69.81 \\
			& TADW & 82.55 & 83.69 & 84.04 & 84.33 & 84.60 & 84.28 & 84.75 & 84.51 & 84.54 \\
			& SNE & 80.01 & 81.37 & 81.70 & 81.82 & 82.08 & 82.04 & 82.22 & 82.23 & 82.42 \\
			& MVC-DNE & 81.12 & 81.86 & 82.12 & 82.33 & 82.54 & 82.36 & 82.37 & 82.21 & 82.37 \\
			& attri2vec-linear & 84.09 & 85.44 & 85.85 & 85.89 & 86.25 & 86.08 & 86.50 & 86.34 & 86.11 \\
			& attri2vec-ReLU & \underline{85.67} & \textbf{86.71} & \textbf{87.07} & \textbf{87.31} & \textbf{87.35} & \textbf{87.46} & \textbf{87.55} & \textbf{87.22} & \textbf{87.55} \\
			& attri2vec-kernel & 84.30 & 85.00 & 85.28 & 85.30 & 85.49 & 85.37 & 85.68 & 85.47 & 85.57 \\
			& attri2vec-sigmoid & \textbf{85.79} & \underline{86.42} & \underline{86.65} & \underline{86.83} & \underline{86.82} & \underline{86.76} & \underline{87.05} & \underline{86.81} & \underline{86.85} \\
			\hline
	\end{tabular}}
	\label{Res_Classification_PubMed}
\end{table*}

\begin{table*}[t]
	\centering
		\caption{Node Classification Results on Facebook}
		\tabcolsep 3pt
		\renewcommand{\arraystretch}{1.1}
		\scalebox{0.85}{
			\begin{tabular}{clcccccccccc}
				\hline
				&Training Ratio & 10\% & 20\% & 30\% & 40\% & 50\% & 60\% & 70\% & 80\% & 90\% \\ \hline
				\multirow{11}{*}{Micro-$F_{1}$(\%)}& DeepWalk & 46.75 & 49.21 & 51.27 & 51.58 & 52.21 & 52.06 & 52.35 & 52.84 & 50.82 \\
				& LINE-1 & 42.43 & 46.62 & 49.13 & 50.38 & 51.28 & 51.41 & 52.11 & 51.98 & 51.32 \\
				& LINE-2 & 37.04 & 43.27 & 46.94 & 48.62 & 49.45 & 49.29 & 49.90 & 49.79 & 50.82 \\
				& SDNE & 38.87 & 45.57 & 48.72 & 50.59 & 51.38 & 51.36 & 51.51 & 52.26 & 51.14 \\
				& TADW & 53.95 & 56.66 & 57.56 & 58.04 & 58.66 & 58.06 & 58.09 & 59.26 & 57.84 \\
				& SNE & 51.32 & 60.15 & 64.01 & 65.26 & 65.95 & 66.93 & 66.16 & 66.85 & 66.13 \\
				& MVC-DNE & 58.12 & 64.80 & \underline{68.17} & \underline{69.87} & \underline{70.51} & \underline{70.37} & \underline{71.04} & \underline{71.34} & \underline{70.92} \\
				& attri2vec-linear & 59.36 & 64.12 & 66.79 & 68.23 & 69.00 & 69.42 & 69.79 & 69.80 & 69.21 \\
				& attri2vec-ReLU & \underline{59.69} & 64.25 & 67.34 & 68.70 & 69.42 & 69.43 & 70.07 & 70.15 & 69.98 \\
				& attri2vec-kernel & \textbf{63.04} & \underline{65.86} & 67.57 & 68.48 & 68.59 & 69.24 & 69.32 & 69.68 & 69.45 \\
				& attri2vec-sigmoid & \textbf{63.06} & \textbf{67.35} & \textbf{69.86} & \textbf{71.38} & \textbf{71.30} & \textbf{72.19} & \textbf{72.19} & \textbf{72.49} & \textbf{72.11} \\
				\hline
				\multirow{11}{*}{Macro-$F_{1}$(\%)} & DeepWalk & 29.85 & 31.04 & 30.55 & 30.83 & 30.76 & 30.31 & 30.15 & 29.76 & 29.29 \\
				& LINE-1 & 29.84 & 31.09 & 31.56 & 31.54 & 31.56 & 31.09 & 31.41 & 30.63 & 30.26 \\
				& LINE-2 & 26.80 & 28.18 & 27.46 & 27.60 & 27.22 & 26.22 & 26.26 & 26.02 & 27.12 \\
				& SDNE & 28.36 & 30.24 & 29.69 & 30.12 & 29.97 & 29.25 & 29.22 & 29.44 & 28.62 \\
				& TADW & 32.17 & 35.80 & 36.72 & 37.29 & 37.94 & 37.23 & 36.85 & 38.29 & 36.97 \\
				& SNE & 40.15 & 43.51 & 43.74 & 43.61 & 43.76 & 43.72 & 42.35 & 42.51 & 41.89 \\
				& MVC-DNE & 45.15 & \underline{47.26} & 48.06 & 47.67 & 47.73 & 47.26 & 47.34 & 47.50 & 47.16 \\
				& attri2vec-linear & 45.31 & 47.07 & 47.72 & 47.04 & 47.25 & 46.66 & 46.72 & 46.50 & 45.50 \\
				& attri2vec-ReLU & \underline{45.47} & 47.00 & \underline{48.27} & \underline{48.26} & \underline{48.13} & \underline{47.88} & \underline{48.17} & \underline{48.08} & \underline{47.39} \\
				& attri2vec-kernel & 44.91 & 46.82 & 46.73 & 46.61 & 46.31 & 46.39 & 45.61 & 45.92 & 45.51 \\
				& attri2vec-sigmoid & \textbf{48.60} & \textbf{50.91} & \textbf{50.99} & \textbf{51.43} & \textbf{50.81} & \textbf{51.43} & \textbf{50.29} & \textbf{50.66} & \textbf{50.73} \\
				\hline
		\end{tabular}}
		\label{Res_Classification_Facebook}
\end{table*}

\begin{table*}[t]
	\centering
		\caption{Node Classification Results on Flickr}
		\tabcolsep 3pt
		\renewcommand{\arraystretch}{1.1}
		\scalebox{0.85}{
			\begin{tabular}{clcccccccccc}
				\hline
				&Training Ratio & 10\% & 20\% & 30\% & 40\% & 50\% & 60\% & 70\% & 80\% & 90\% \\ \hline
				\multirow{11}{*}{Micro-$F_{1}$(\%)}& DeepWalk & 40.25 & 45.60 & 47.67 & 49.05 & 49.74 & 50.30 & 50.88 & 51.76 & 51.72 \\
				& LINE-1 & 42.97 & 51.43 & 55.07 & 57.23 & 58.30 & 58.88 & 59.69 & 60.67 & 59.91 \\
				& LINE-2 & 39.01 & 48.51 & 52.31 & 54.01 & 55.14 & 56.40 & 56.11 & 56.44 & 56.53 \\
				& SDNE & 38.80 & 47.63 & 50.50 & 51.92 & 52.93 & 53.48 & 54.21 & 54.82 & 54.32 \\
				& TADW & 62.17 & 65.17 & 66.32 & 67.11 & 67.17 & 67.97 & 68.12 & 68.16 & 67.79 \\
				& SNE & 40.70 & 49.73 & 53.42 & 55.53 & 56.12 & 57.66 & 57.38 & 58.46 & 58.34 \\
				& MVC-DNE & 56.33 & 64.98 & 68.54 & 70.22 & 70.77 & 72.18 & 71.90 & 72.69 & 72.56 \\
				& attri2vec-linear & 62.04 & 68.95 & 71.99 & 73.28 & 74.31 & 75.04 & 75.04 & 75.77 & 75.85 \\
				& attri2vec-ReLU & 61.27 & 69.12 & 72.14 & 73.95 & 74.71 & 75.38 & 75.66 & 75.99 & 76.39 \\
				& attri2vec-kernel & \underline{75.20} & \underline{77.05} & \underline{77.33} & \underline{78.04} & \underline{78.31} & \underline{78.77} & \underline{78.78} & \underline{79.65} & \underline{78.82} \\
				& attri2vec-sigmoid & \textbf{75.40} & \textbf{78.25} & \textbf{79.38} & \textbf{79.98} & \textbf{80.41} & \textbf{80.96} & \textbf{80.74} & \textbf{81.66} & \textbf{80.18} \\
				\hline
				\multirow{11}{*}{Macro-$F_{1}$(\%)} & DeepWalk & 39.67 & 44.65 & 46.48 & 47.80 & 48.43 & 48.87 & 49.48 & 50.13 & 49.99 \\
				& LINE-1 & 42.89 & 51.00 & 54.41 & 56.50 & 57.63 & 57.99 & 58.87 & 59.65 & 58.92 \\
				& LINE-2 & 38.99 & 47.94 & 51.55 & 53.15 & 54.19 & 55.22 & 55.03 & 55.13 & 55.01 \\
				& SDNE & 38.62 & 47.12 & 49.79 & 51.14 & 52.09 & 52.55 & 53.29 & 53.73 & 53.17 \\
				& TADW & 61.54 & 64.52 & 65.90 & 66.71 & 66.69 & 67.36 & 67.60 & 67.59 & 67.20 \\
				& SNE & 40.55 & 49.11 & 52.57 & 54.53 & 55.12 & 56.43 & 56.23 & 57.19 & 57.10 \\
				& MVC-DNE & 56.20 & 64.78 & 68.32 & 69.97 & 70.57 & 71.88 & 71.65 & 72.38 & 72.19 \\
				& attri2vec-linear & 61.93 & 68.77 & 71.77 & 73.00 & 74.10 & 74.70 & 74.74 & 75.43 & 75.38 \\
				& attri2vec-ReLU & 61.27 & 68.97 & 71.94 & 73.69 & 74.50 & 75.09 & 75.40 & 75.65 & 75.92 \\
				& attri2vec-kernel & \underline{74.87} & \underline{76.81} & \underline{77.12} & \underline{77.77} & \underline{78.11} & \underline{78.53} & \underline{78.57} & \underline{79.36} & \underline{78.48} \\
				& attri2vec-sigmoid & \textbf{75.14} & \textbf{78.02} & \textbf{79.15} & \textbf{79.68} & \textbf{80.21} & \textbf{80.69} & \textbf{80.48} & \textbf{81.33} & \textbf{79.81} \\
				\hline
		\end{tabular}}
		\label{Res_Classification_Flickr}
\end{table*}

To evaluate the quality of node representations learned by different network embedding algorithms, we carry out node classification experiments by taking node representations as features. We randomly select a ratio of training samples and use them to train an SVM classifier (implemented by LIBLINEAR~\citep{fan2008liblinear}), and then test its performance on the held-out samples. We vary the training ratio from 10\% to 90\% with a step of 10\%. For each training ratio, we repeat the random training and test set split for 10 times and report the averaged Micro-$F_{1}$ and Macro-$F_{1}$ values as final results. Tables~\ref{Res_Classification_Citeseer}-\ref{Res_Classification_Flickr} report the node classification results on Citeseer, DBLP, PubMed, Facebook and Flickr. For Micro-$F_{1}$ and Macro-$F_{1}$, the best and second performer is highlighted by \textbf{bold} and \underline{underline}, respectively. 

From Tables \ref{Res_Classification_Citeseer}-\ref{Res_Classification_Flickr}, we can see that attri2vec achieves the best classification performance on all networks with all training ratios, with attri2vec-sigmoid performing best on Citeseer, Facebook and Flick and attri2vec-ReLU performing best on DBLP and PubMed. This suggests that, by discovering a latent node attribute subspace, attri2vec provides the best way to complement network structure with node attributes towards learning node representations. Though attri2vec-ReLU performs best on DBLP and PubMed, the attri2vec-sigmoid achieves the comparable classification performance. Overall, under the attri2vec framework, the non-linear mappings (approximated kernel and sigmoid mapping) outperform the linear mappings (linear and ReLU mapping). Network structure and node content usually exhibit non-linear correlations, especially for social networks. As a result, node attribute transformation with a non-linear mapping provides a better way to fuse network structure into a node attribute subspace.

We can also observe that, in most cases, attributed network embedding algorithms (TADW, SNE, MVC-DNE and attri2vec) significantly outperform the only structure preserving network embedding algorithms (DeepWalk, LINE-1, LINE-2 and SDNE). This proves that node content features are able to provide an essential complement to network structure to learn better node representations. Among the structure preserving network embedding algorithms, DeepWalk achieves the best classification performance. Compared with LINE-1, LINE-2 and SDNE that only capture the local first-order and second-order proximity, DeepWalk takes advantage of random walks to preserve the higher-order proximity, which contributes to DeepWalk's superior performance over LINE-1, LINE-2 and SDNE.

\subsection{Node Clustering Experiments}

\begin{table}[t]
	\centering
	\caption{Node Clustering Results on DBLP}
	\scalebox{1}{
		\begin{tabular}{cccc}
			\hline
			&Accuracy(\%) & Fvalue(\%) & NMI(\%)  \\\hline
			DeepWalk & 70.71 & 64.50 & 36.92 \\
			LINE-1 & 58.02 & 54.57 & 18.88 \\
			LINE-2 & 61.23 & 47.44 & 18.33 \\
			SDNE & 52.33 & 42.61 & 11.73 \\
			TADW & 59.63 & 58.16 & 17.81 \\
			SNE & 61.37 & 49.96 & 16.19 \\
			MVC-DNE & 66.25 & 60.42 & 28.30 \\
			attri2vec-linear & 71.05 & 63.39 & 34.31 \\
			attri2vec-ReLU & \textbf{76.69} & \textbf{69.67} & \textbf{44.34} \\
			attri2vec-kernel & 73.18 & 67.96 & 39.87 \\
			attri2vec-sigmoid & 72.07 & 66.08 & 38.26 \\
			\hline
	\end{tabular}}
	\label{Res_Clustering_DBLP}
\end{table}

\begin{table}[t]
	\centering
	\caption{Node Clustering Results on Facebook}
	\scalebox{1}{
		\begin{tabular}{cccc}
			\hline
			&Accuracy(\%) & Fvalue(\%) & NMI(\%)  \\\hline
			DeepWalk & 52.65 & 39.10 & 2.03 \\
			LINE-1 & 51.91 & 46.27 & 1.34 \\
			LINE-2 & 51.84 & 38.97 & 1.21 \\
			SDNE & 51.88 & 42.75 & 2.11 \\
			TADW & 51.84 & 46.23 & 7.76 \\
			SNE & 54.12 & 39.28 & 9.10 \\
			MVC-DNE & 52.39 & 38.05 & 5.56 \\
			attri2vec-linear & 53.89 & 41.62 & 6.52 \\
			attri2vec-ReLU & 53.66 & 41.74 & 5.70 \\
			attri2vec-kernel & \textbf{66.77} & \textbf{50.32} & \textbf{17.15} \\
			attri2vec-sigmoid & 53.26 & 41.71 & 5.55 \\
			\hline
	\end{tabular}}
	\label{Res_Clustering_Facebook}
\end{table}

To make further comparisons between the proposed attri2vec algorithm and baseline methods, on DBLP and Facebook, we also conduct node clustering experiments. We apply $K$-means algorithm to the learned node representations, and partition network nodes into 4 groups for DBLP and Facebook and use node labels as clustering ground truth. To reduce the variance caused by random initialization, we repeat $K$-means clustering for 20 times and report the averaged Accuracy, Fvalue and NMI (normalized mutual information~\citep{strehl2002cluster}). Tables \ref{Res_Clustering_DBLP}-\ref{Res_Clustering_Facebook} give the clustering results on DBLP and Facebook. As is shown in Tables \ref{Res_Clustering_DBLP}-\ref{Res_Clustering_Facebook}, the proposed attri2vec algorithm achieves the best clustering performance, which further proves its advantage in learning informative node representations for attributed networks over the state-of-the-art baselines.

\subsection{Comparison of Gradient Descent $vs.$ Stochastic Gradient Descent }

In this section, we choose the Facebook network as a case study to compare the efficacy and efficiency of two difference optimization strategies: stochastic gradient descent (SGD) used by attri2vec and Gradient Descent (GD) used by UPP-SNE~\citep{zhang2017user}, for solving the optimization problem Eq.~(\ref{optimization_prob}). We compare UPP-SNE with attri2vec-kernel, as they both use the approximated kernel mapping to construct node representations. To make a fair comparison, for UPP-SNE, we set the number of iterations for gradient descent to 40, and for attri2vec, we set the number of iteration as $nnz(n(v_{i},v_{j}))\times40$, where $nnz(n(v_{i},v_{j}))$ denotes the number of non-zero values of $n(v_{i},v_{j})$. 

Table~\ref{Facebook_GD_SGD} compares both node classification performance and training time of UPP-SNE and attri2vec-kernel. Here, we report the classification results with a training ratio of 50\%. As can be seen, attri2vec-kernel with SGD offers performance gains with an increase of 11.60\% in Micro-F1, and 30.33\% in Macro-F1, respectively, as SGD alleviates the local minima problem. On the other hand, SGD significantly reduces the training time of GD, with a decrease of 55.36\%. This demonstrates the advantage of attri2vec over UPP-SNE in terms of both effectiveness and efficiency. 

\begin{table*}[t]
	\centering
	\caption{Performance Comparison between GD and SGD on Facebook}
	\begin{tabular}{cccc}
		\hline
		Method & Micro-$F_{1}$ & Macro-$F_{1}$ & Training Time (s)\\\hline
		UPP-SNE (GD) & 0.6192 &  0.3535 & 24515.89 \\
		attri2vec-kernel (SGD) & 0.6910 & 0.4607 & 10944.35 \\
		Gain & 11.60\% $\uparrow$ & 30.33\% $\uparrow$ & 55.36\% $\downarrow$ \\\hline
	\end{tabular}
	\label{Facebook_GD_SGD}
\end{table*}

\subsection{Comparison of Shallow $vs.$ Deep Mapping}

In this section, we select Citeseer and DBLP to study the performance change of attri2vec when using a deep neural network rather than a shallow one-layer non-linear mapping to construct node representations from node features. Here, we construct node representations through a neural network with 2 hidden layers, with the number of neurons at each layer set to 3,703-256-128 and 2,476-256-128 on Citeseer and DBLP respectively, and denote this method as attri2vec-deep. We compare attri2vec-deep with attri2vec-sigmoid. For both methods, we set the number of iterations as 100 million.

Tables~\ref{citeseer_deep_shallow}-\ref{DBLP_deep_shallow} compare the classification performance and the training time of attri2vec-linear and attri2vec-deep on Citeseer and DBLP. Again, results with a training ratio of 50\% are reported. As can be seen, attri2vec-sigmoid is significantly more efficient than attri2vec-deep for training. From Table~\ref{DBLP_deep_shallow}, attri2vec-deep can be seen to perform slightly better than attri2vec-sigmoid on DBLP, which demonstrates the advantage of deep neural network in characterizing the complex relations between network structure and node content. However, on Citeseer, attri2vec-deep achieves even worse classification performance than attri2vec-sigmoid. This might be attributed to two reasons: (1) attri2vec-deep has a lot more parameters, which requires more iterations to properly fitting the parameters to obtain the best models; (2) compared with attri2vec-sigmoid, attri2vec-deep is prone to the local minima, which requires more advanced SGD techniques to obtain better solutions. In summary, compared with the deep-neural network, the one-layer non-linear mapping is good enough to learn reasonably high-quality node representations with much less computational cost. 

\begin{table*}[t]
	\centering
	\caption{Performance Comparison of Shallow and Deep Mapping on Citeseer}
	\begin{tabular}{cccc}
		\hline
		Method & Micro-$F_{1}$ & Macro-$F_{1}$ & Training Time (s)\\\hline
		attri2vec-sigmoid & 0.7027 &  0.6569 & 29405.51 \\
		attri2vec-deep & 0.6630 & 0.6076 & 380331.50 \\
		Gain & 5.65\% $\downarrow$ & 7.50\% $\downarrow$ & 1193.40\% $\uparrow$ \\\hline
	\end{tabular}
	\label{citeseer_deep_shallow}
\end{table*}

\begin{table*}[t]
	\centering
	\caption{Performance Comparison of Shallow and Deep Mapping on DBLP}
	\begin{tabular}{cccc}
		\hline
		Method & Micro-$F_{1}$(\%) & Macro-$F_{1}$(\%) & Training Time (s)\\\hline
		attri2vec-sigmoid & 82.52 &  76.88 & 5310.19 \\
		attri2vec-deep & 83.72 & 78.29 & 319881.87 \\
		Gain & 1.45\% $\uparrow$ & 1.83\% $\uparrow$ & 5923.96\% $\uparrow$ \\\hline
	\end{tabular}
	\label{DBLP_deep_shallow}
\end{table*}


\subsection{Experiments on Out-of-sample Extension}

In this section, we study the performance of attri2vec in solving the out-of-sample problem, which is achieved by constructing representations for new coming nodes from their content attributes through the learned mapping function. We compare attri2vec with two baselines (1) MVC-DNE~\citep{yang2017properties}, which is also able to infer representations for new coming nodes using node attributes, and (2) node content features that construct node representations without mapping functions. For the three methods, we consistently set the dimension of out-of-sample node representations to 128. Singular Value Decomposition (SVD) is used to do dimension reduction on node content features.

\begin{table*}[t]
	\centering
	\caption{A Summary of Time Stamped DBLP Subgraphs}
	\begin{tabular}{cccc}
		\hline
		Subgraph & \# of nodes & \# of edges & \# of out-of-sample nodes\\\hline
		DBLP2006 & 12,820 & 28,809 & 5,628 \\
		DBLP2007 & 14,286 & 32,336 & 4,162 \\
		DBLP2008 & 15,745 & 36,907 & 2,703 \\
		DBLP2009 & 16,960 & 40,460 & 1,488 \\\hline
	\end{tabular}
	\label{DBLP_subgraph_summary}
\end{table*}

We select the DBLP network to conduct the experiments, where each node has a time stamp. From the DBLP network, we construct four subgraphs using papers published before 2006, 2007, 2008 and 2009, and denote the four subgraphs as DBLP2006, DBLP2007, DBLP2008, and DBLP2009. For each subgraph, the remaining papers are taken as out-of-sample nodes. The statistics of the four DBLP subgraphs and their corresponding out-of-sample nodes are given in Table \ref{DBLP_subgraph_summary}. To evaluate the quality of the learned out-of-sample node representations, we conduct node classification and link prediction experiments. 

\begin{table*}
	\centering
	\tabcolsep 4pt
	\caption{Node Classification Results for Out-of-sample nodes on DBLP}
	\scalebox{1.0}{
		\begin{tabular}{cccccc}
			\hline
			& method & DBLP2006 & DBLP2007 & DBLP2008 & DBLP2009\\\hline
			\multirow{3}{*}{Micro-$F_{1}$(\%)} & feature & 65.38 & 64.01 & 61.75 & 62.12\\
			& MVC-DNE & 64.96 & 62.74 & 61.45 & 61.68 \\
			& attri2vec-sigmoid & \textbf{67.66} & \textbf{66.79} & \textbf{64.87} & \textbf{67.05} \\\hline
			\multirow{3}{*}{Macro-$F_{1}$(\%)} &  feature & 60.49 & 59.77 & 58.22 & 57.14 \\
			& MVC-DNE & 59.53 & 57.34 & 56.96 & 55.12 \\
			& attri2vec-sigmoid & \textbf{63.94} & \textbf{63.69} & \textbf{62.35} & \textbf{62.50} \\\hline
	\end{tabular}}
	\label{DBLP_classification_outofsample}
\end{table*}

\begin{table*}[t]
	\centering
	\caption{Operators to Construct Edge Features}
	\scalebox{1.0}{
		\begin{tabular}{ccc}
			\hline
			Operator & Symbol & Definition\\\hline
			Average & $\boxplus$ & $\left[\mathrm{\Phi}(v_{i})\boxplus\mathrm{\Phi}(v_{j})\right]_{k}=\frac{\mathrm{\Phi}_{k}(v_{i})+\mathrm{\Phi}_{k}(v_{j})}{2}$ \\
			Hadamard & $\boxdot$ & $\left[\mathrm{\Phi}(v_{i})\boxdot\mathrm{\Phi}(v_{j})\right]_{k}=\mathrm{\Phi}_{k}(v_{i})\cdot\mathrm{\Phi}_{k}(v_{j})$\\
			Weighted-L1 & ${\lVert\cdot\rVert}_{\bar{1}}$ & ${\lVert\mathrm{\Phi}(v_{i})\cdot\mathrm{\Phi}(v_{j})\rVert}_{\bar{1}k}=|\mathrm{\Phi}_{k}(v_{i})-\mathrm{\Phi}_{k}(v_{j})|$\\
			Weighted-L2 & ${\lVert\cdot\rVert}_{\bar{2}}$ & ${\lVert\mathrm{\Phi}(v_{i})\cdot\mathrm{\Phi}(v_{j})\rVert}_{\bar{2}k}=(\mathrm{\Phi}_{k}(v_{i})-\mathrm{\Phi}_{k}(v_{j}))^{2}$\\\hline
	\end{tabular}}
	\label{edge_feature}
\end{table*}

\begin{table*}[t]
	\centering
	\tabcolsep 4pt
	\caption{AUC Values (\%) for Predicting the Links of Out-of-sample Nodes on DBLP}
	\scalebox{1.0}{
		\begin{tabular}{cccccc}
			\hline
			method & operator & DBLP2006 & DBLP2007 & DBLP2008 & DBLP2009 \\\hline
			\multirow{4}{*}{feature} & Average & 53.31 & 53.52 & 54.39 & 54.44 \\
			 & Hadamard & 76.07 & 76.43 & 76.17 & 76.38 \\
			 & Weighted-L1 & 64.84 & 65.84 & 65.60 & 65.42 \\
			 & Weighted-L2 & 65.20 & 66.17 & 65.92 & 65.74 \\\hline
			\multirow{4}{*}{MVC-DNE} & Average & 51.78 & 52.74 & 54.18 & 55.05 \\
			& Hadamard & 51.07 & 51.12 & 51.58 & 51.92 \\
			& Weighted-L1 & 49.70 & 50.94 & 50.69 & 49.19 \\
			& Weighted-L2 & 49.95 & 51.11 & 50.80 & 49.30 \\\hline
			\multirow{4}{*}{attri2vec-sigmoid} & Average & 53.41 & 54.22 & 56.00 & 57.01 \\
			& Hadamard & 80.82 & 81.98 & 82.56 & 83.88 \\
			& Weighted-L1 & 87.99 & 89.16 & 90.43 & 91.38 \\
			& Weighted-L2 & \textbf{88.11} & \textbf{89.33} & \textbf{90.66} & \textbf{91.70} \\\hline
	\end{tabular}}
	\label{DBLP_linkprediction_outofsample}
\end{table*}

For node classification, we first train an SVM classifier (with the LIBLINEAR implementation~\citep{fan2008liblinear}) on the randomly selected 50\% learned in-sample node representations, and then apply the learned SVM classifier to the out-of-sample node representations. To reduce the variance caused by random training sample selection, we repeat the training and test process for 10 times and report the averaged Micro-$F_{1}$ and Macro-$F_{1}$. Table~\ref{DBLP_classification_outofsample} gives the node classification results for out-of-sample nodes. We can see that attri2vec yields the best classification performance. This demonstrates the effectiveness of attri2vec in solving the out-of-sample problem.

To carry out link prediction experiments, following ~\citep{grover2016node2vec}, we construct edge features from node representations with the operators given in Table~\ref{edge_feature}. To generate the training set, on the in-sample subgraphs, for each connected node pair $(v_{i},v_{j})$, we randomly sample a negative node pair $(v_{i},v_{k})$ with no edges observed between them in the current time stamp. Similarly, to construct test set, for each edge connecting to the out-of-sample nodes, we randomly sample a negative node pair, with no ground-truth edges between them. On the generated edge features, we adopt SVM to perform training and testing. Table~\ref{DBLP_linkprediction_outofsample} gives the AUC values of different methods for predicting the links of out-sample-nodes. As shown in the table, attri2vec-sigmoid with the Weighted-L2 operator performs best in predicting links for new coming out-of-sample nodes. This proves the potential of attri2vec in accurately recommending links for new coming out-of-sample nodes through the inferred node representations.

\begin{figure*}[t]
	\centering
	\includegraphics[width=3.5in]{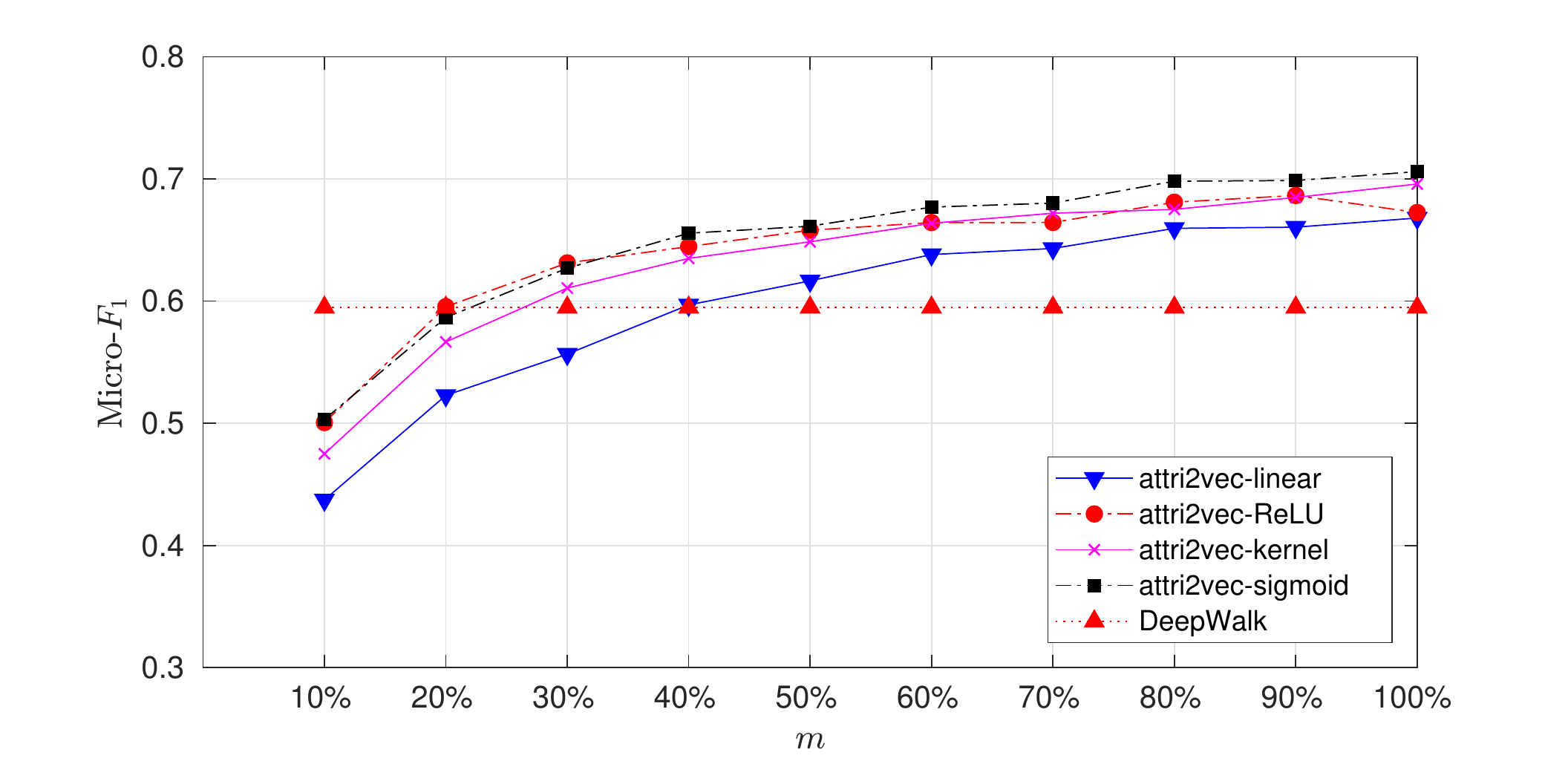}
	\caption{The performance of attri2vec with varying attribute dimensions.}
	\label{fig:attribute_dim} 
\end{figure*}

\subsection{Experiments on Attribute Sparsity}
To alleviate the data sparsity in node attributes, the proposed attri2vec algorithm tries to discover a structure preserving attribute subspace by performing a series of linear or nonlinear mappings on node attributes. In this section, we conduct experiments to study the ability of attri2vec in handling varying extent of attribute sparsity. We choose the Citeseer network as a case study and randomly select 10\% to 100\% attribute dimensions to learn node representations. Fig.~\ref{fig:attribute_dim} plots the performance of attri2vec together with DeepWalk as a baseline with varying attribute dimensions. The performance is measured by the node classification Micro-$F_{1}$ value with 50\% training ratio. As shown in Fig.~\ref{fig:attribute_dim}, the attri2vec algorithm is relatively robust to attribute sparsity: attri2vec-ReLU, attri2vec-kernel and attri2vec-sigmoid start to outperform DeepWalk with 30\% attribute dimension, and the performance of attri2vec variants is relatively stable when attribute dimensions are in the range of 50\% and 100\%. 

\subsection{Parameter Sensitivity Study}

In this section, we report the sensitivity study of the proposed attri2vec algorithm by analyzing its detailed performance on the DBLP network. In the experiments, we consider three important parameters: the maximum number of iterations, random walk window size $t$, and embedding dimension $d$. In order to draw conclusive observations, we in turn fix any two of the three parameters and investigate the performance change of attri2vec when the remaining one varies. Micro-$F_{1}$ of node classification with 50\% training ratio is used to evaluate the performance. 

\begin{figure*}[h]
	\centering
	\subfigure[$\#iterations$]{
		\label{fig:para:subfig:iteration}
		\includegraphics[width=3.5in]{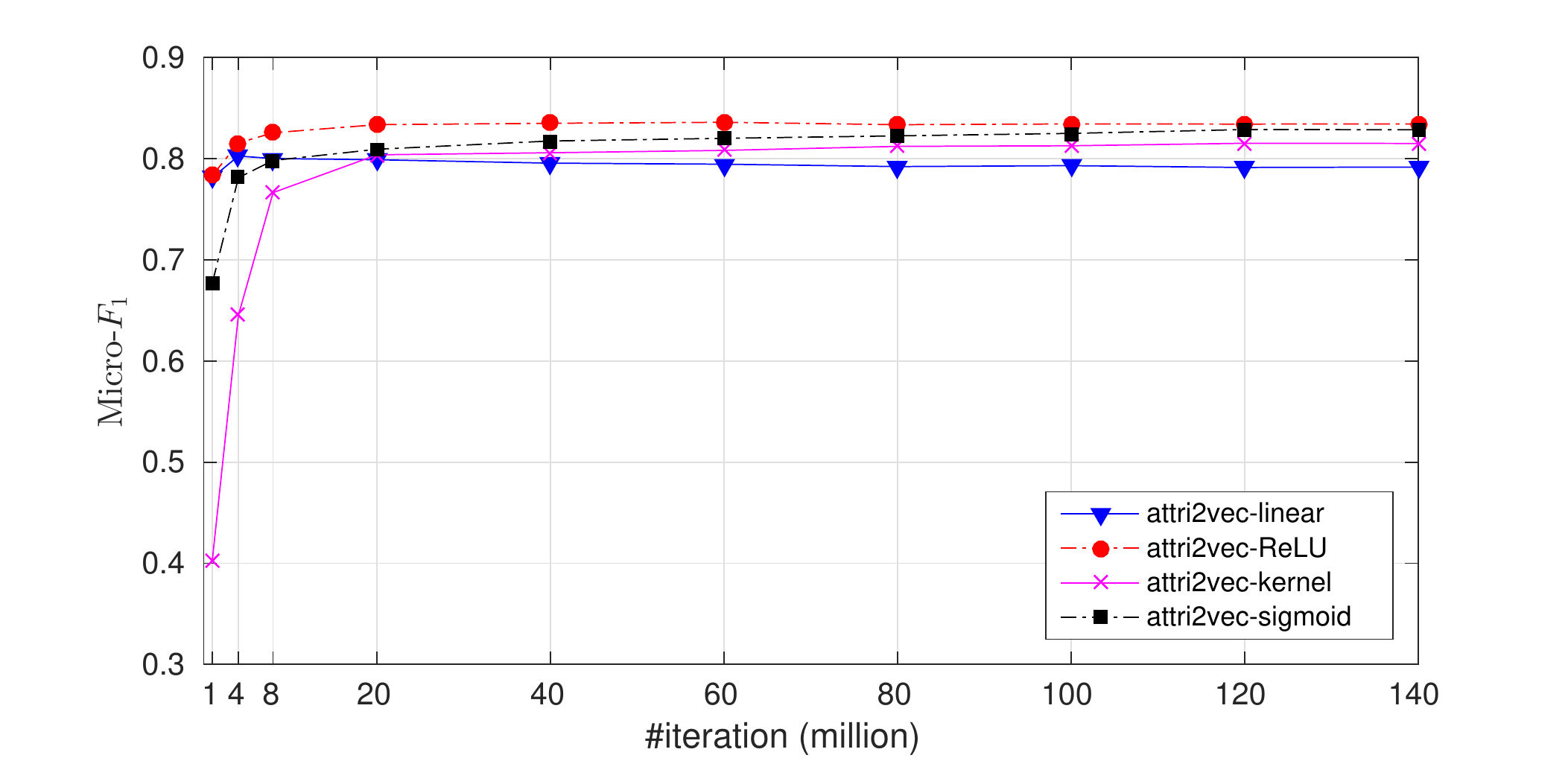}}
	\subfigure[$t$]{
		\label{fig:para:subfig:window}
		\includegraphics[width=3.5in]{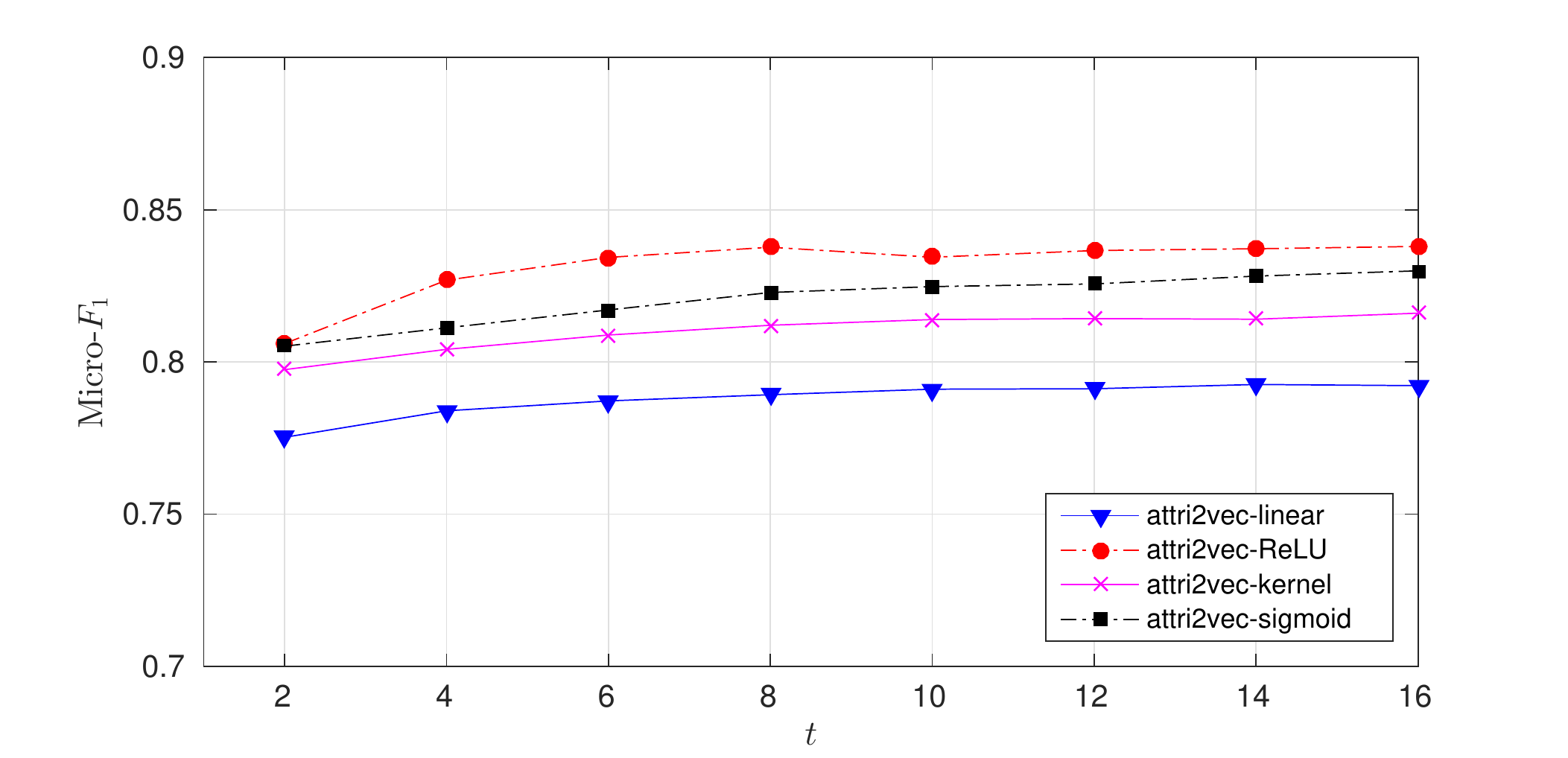}}
	\subfigure[$d$]{
		\label{fig:para:subfig:dim}
		\includegraphics[width=3.5in]{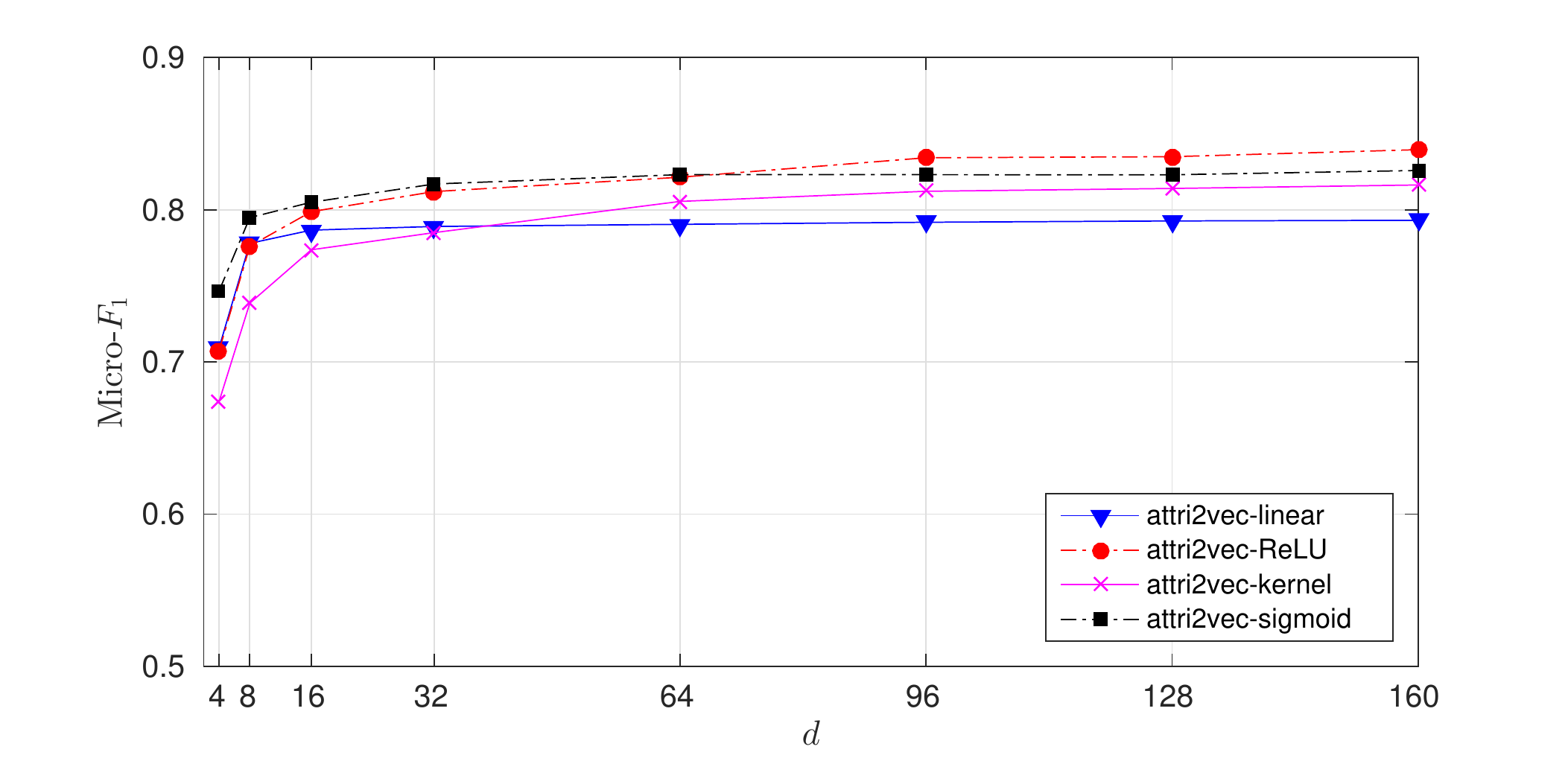}}
	\caption{Parameter sensitivity study of the algorithm performance (Micro-$F_1$) in terms of (a): the maximum number of iterations, (b): the window size of the random walks $t$, and (c): embedding dimension $d$.  }
	\label{fig:para} 
\end{figure*}

Fig.~\ref{fig:para} shows the parameter sensitivity experimental results. We can find that as the parameters increase, classification accuracy of attri2vec  gradually increases and stabilizes after reaching a threshold.

\section{Conclusion}
\label{sec:conclusion}

In this paper, we proposed a unified framework for attributed network embedding, attri2vec, that learns network node representations by discovering a latent node attribute subspace via a network structure guided transformation on the original attribute space. We argued that, because network structure and node attributes have different feature spaces, the data distributions of two distinct features spaces might be inconsistent, making the existing attributed network embedding algorithms fail to achieve satisfactory performance. This motivated us to find a latent attribute subspace, which can respect network structure in a more consistent manner to obtain high-quality node representations. By performing a series of linear and non-linear mappings on node attributes, attri2vec embeds network nodes into a structure preserving attribute subspace. To preserve the network structure, DeepWalk~\citep{perozzi2014deepwalk} mechanism is employed, which makes nodes sharing similar neighbors embedded closely in the attribute subspace. In this way, network structure and node content features are integrated together seamlessly to obtain informative node representations. To learn node embeddings efficiently, we developed an online stochastic gradient descent algorithm to solve the formulated optimization problem, which improves both learning effectiveness and efficiency. As an additional advantage, attri2vec also provides a potential solution to solve the out-of-sample problem, where representations of new coming nodes can be learned using their available node attributes via the learned mapping function. Experiments of node classification and node clustering on five real-world networks verify the effectiveness and efficiency of attri2vec. We also study the ability of attri2vec to solve the out-of-sample problem, which yields promising results.

\begin{acknowledgements}
The work is supported by the US National Science Foundation (NSF) through grant IIS-1763452, and the Australian Research Council (ARC) through grant LP160100630 and DP180100966. Daokun Zhang is supported by China Scholarship Council (CSC) with No. 201506300082 and a supplementary postgraduate scholarship from CSIRO.

\end{acknowledgements}

\bibliographystyle{spbasic} 
\bibliography{attri2vec-bibliography}

\begin{thebibliography}{36}
\providecommand{\natexlab}[1]{#1}
\providecommand{\url}[1]{{#1}}
\providecommand{\urlprefix}{URL }
\expandafter\ifx\csname urlstyle\endcsname\relax
  \providecommand{\doi}[1]{DOI~\discretionary{}{}{}#1}\else
  \providecommand{\doi}{DOI~\discretionary{}{}{}\begingroup
  \urlstyle{rm}\Url}\fi
\providecommand{\eprint}[2][]{\url{#2}}

\bibitem[{Bianconi et~al.(2009)Bianconi, Pin, and
  Marsili}]{Bianconi2009Assessing}
Bianconi G, Pin P, Marsili M (2009) Assessing the relevance of node features
  for network structure. Proceedings of the National Academy of Sciences
  106(28):11433--11438

\bibitem[{Cao et~al.(2015)Cao, Lu, and Xu}]{cao2015grarep}
Cao S, Lu W, Xu Q (2015) {GraRep}: Learning graph representations with global
  structural information. In: Proceedings of the 24th ACM International
  Conference on Information and Knowledge Management, ACM, pp 891--900

\bibitem[{Cao et~al.(2016)Cao, Lu, and Xu}]{cao2016deep}
Cao S, Lu W, Xu Q (2016) Deep neural networks for learning graph
  representations. In: Proceedings of the 30th AAAI Conference on Artificial
  Intelligence, AAAI Press, pp 1145--1152

\bibitem[{Fan et~al.(2008)Fan, Chang, Hsieh, Wang, and Lin}]{fan2008liblinear}
Fan RE, Chang KW, Hsieh CJ, Wang XR, Lin CJ (2008) {LIBLINEAR}: A library for
  large linear classification. Journal of Machine Learning Research
  9(Aug):1871--1874

\bibitem[{Grover and Leskovec(2016)}]{grover2016node2vec}
Grover A, Leskovec J (2016) node2vec: Scalable feature learning for networks.
  In: Proceedings of the 22nd ACM SIGKDD International Conference on Knowledge
  Discovery and Data Mining, ACM, pp 855--864

\bibitem[{Guo et~al.(2018)Guo, Pan, Zhu, and Zhang}]{guo2018cfond}
Guo T, Pan S, Zhu X, Zhang C (2018) {CFOND}: Consensus factorization for
  co-clustering networked data. IEEE Transactions on Knowledge and Data
  Engineering

\bibitem[{Gutmann and Hyv{\"a}rinen(2012)}]{gutmann2012noise}
Gutmann MU, Hyv{\"a}rinen A (2012) Noise-contrastive estimation of unnormalized
  statistical models, with applications to natural image statistics. Journal of
  Machine Learning Research 13(Feb):307--361

\bibitem[{Hamilton et~al.(2017)Hamilton, Ying, and
  Leskovec}]{hamilton2017inductive}
Hamilton W, Ying Z, Leskovec J (2017) Inductive representation learning on
  large graphs. In: Advances in Neural Information Processing Systems, pp
  1024--1034

\bibitem[{Hotelling(1936)}]{hotelling1936relations}
Hotelling H (1936) Relations between two sets of variates. Biometrika
  28(3/4):321--377

\bibitem[{Huang et~al.(2017{\natexlab{a}})Huang, Li, and
  Hu}]{huang2017accelerated}
Huang X, Li J, Hu X (2017{\natexlab{a}}) Accelerated attributed network
  embedding. In: Proceedings of the 2017 SIAM International Conference on Data
  Mining, SIAM, pp 633--641

\bibitem[{Huang et~al.(2017{\natexlab{b}})Huang, Li, and Hu}]{huang2017label}
Huang X, Li J, Hu X (2017{\natexlab{b}}) Label informed attributed network
  embedding. In: Proceedings of the 10th ACM International Conference on Web
  Search and Data Mining, ACM, pp 731--739

\bibitem[{Kuang et~al.(2012)Kuang, Ding, and Park}]{kuang2012symmetric}
Kuang D, Ding C, Park H (2012) Symmetric nonnegative matrix factorization for
  graph clustering. In: Proceedings of the 2012 SIAM international conference
  on data mining, SIAM, pp 106--117

\bibitem[{Leskovec and Mcauley(2012)}]{leskovec2012learning}
Leskovec J, Mcauley JJ (2012) Learning to discover social circles in ego
  networks. In: Advances in neural information processing systems, pp 539--547

\bibitem[{Levy and Goldberg(2014)}]{levy2014neural}
Levy O, Goldberg Y (2014) Neural word embedding as implicit matrix
  factorization. In: Advances in neural information processing systems, pp
  2177--2185

\bibitem[{Li et~al.(2014)Li, Ahmed, Ravi, and Smola}]{li2014reducing}
Li AQ, Ahmed A, Ravi S, Smola AJ (2014) Reducing the sampling complexity of
  topic models. In: Proceedings of the 20th ACM SIGKDD international conference
  on Knowledge discovery and data mining, ACM, pp 891--900

\bibitem[{Li et~al.(2016)Li, Zhu, and Zhang}]{li2016discriminative}
Li J, Zhu J, Zhang B (2016) Discriminative deep random walk for network
  classification. In: Proceedings of the 54th Annual Meeting of the Association
  for Computational Linguistics, vol~1, pp 1004--1013

\bibitem[{Liao et~al.(2018)Liao, He, Zhang, and Chua}]{liao2018attributed}
Liao L, He X, Zhang H, Chua TS (2018) Attributed social network embedding. IEEE
  Transactions on Knowledge and Data Engineering

\bibitem[{Mikolov et~al.(2013)Mikolov, Sutskever, Chen, Corrado, and
  Dean}]{mikolov2013distributed}
Mikolov T, Sutskever I, Chen K, Corrado GS, Dean J (2013) Distributed
  representations of words and phrases and their compositionality. In: Advances
  in Neural Information Processing Systems, pp 3111--3119

\bibitem[{Natarajan and Dhillon(2014)}]{natarajan2014inductive}
Natarajan N, Dhillon IS (2014) Inductive matrix completion for predicting
  gene--disease associations. Bioinformatics 30(12):i60--i68

\bibitem[{Newman(2006)}]{newman2006finding}
Newman ME (2006) Finding community structure in networks using the eigenvectors
  of matrices. Physical review E 74(3):036104

\bibitem[{Pan et~al.(2016)Pan, Wu, Zhu, Zhang, and Wang}]{pan2016tri}
Pan S, Wu J, Zhu X, Zhang C, Wang Y (2016) Tri-party deep network
  representation. In: Proceedings of the 25th International Joint Conference on
  Artificial Intelligence, pp 1895--1901

\bibitem[{Perozzi et~al.(2014)Perozzi, Al-Rfou, and
  Skiena}]{perozzi2014deepwalk}
Perozzi B, Al-Rfou R, Skiena S (2014) {DeepWalk}: Online learning of social
  representations. In: Proceedings of the 20th ACM SIGKDD International
  Conference on Knowledge Discovery and Data Mining, ACM, pp 701--710

\bibitem[{Rahimi and Recht(2008)}]{rahimi2008random}
Rahimi A, Recht B (2008) Random features for large-scale kernel machines. In:
  Advances in neural information processing systems, pp 1177--1184

\bibitem[{Reagans and McEvily(2003)}]{reagans2003network}
Reagans R, McEvily B (2003) Network structure and knowledge transfer: The
  effects of cohesion and range. Administrative Science Quarterly 48(2)

\bibitem[{Strehl and Ghosh(2002)}]{strehl2002cluster}
Strehl A, Ghosh J (2002) Cluster ensembles---a knowledge reuse framework for
  combining multiple partitions. Journal of machine learning research
  3(Dec):583--617

\bibitem[{Subbaraj and Sundan(2015)}]{subbarai2015what}
Subbaraj K, Sundan B (2015) What happens next? prediction of disastrous links
  in covert networks. Disaster Advances 8:53--60

\bibitem[{Tang et~al.(2015)Tang, Qu, Wang, Zhang, Yan, and Mei}]{tang2015line}
Tang J, Qu M, Wang M, Zhang M, Yan J, Mei Q (2015) {LINE}: Large-scale
  information network embedding. In: Proceedings of the 24th International
  Conference on World Wide Web, ACM, pp 1067--1077

\bibitem[{Vincent et~al.(2010)Vincent, Larochelle, Lajoie, Bengio, and
  Manzagol}]{vincent2010stacked}
Vincent P, Larochelle H, Lajoie I, Bengio Y, Manzagol PA (2010) Stacked
  denoising autoencoders: Learning useful representations in a deep network
  with a local denoising criterion. Journal of Machine Learning Research
  11(Dec):3371--3408

\bibitem[{Wang et~al.(2016)Wang, Cui, and Zhu}]{wang2016structural}
Wang D, Cui P, Zhu W (2016) Structural deep network embedding. In: Proceedings
  of the 22nd ACM SIGKDD International Conference on Knowledge Discovery and
  Data Mining, ACM, pp 1225--1234

\bibitem[{Wang et~al.(2017)Wang, Cui, Wang, Pei, Zhu, and
  Yang}]{wang2017community}
Wang X, Cui P, Wang J, Pei J, Zhu W, Yang S (2017) Community preserving network
  embedding. In: Proceedings of the 31st AAAI Conference on Artificial
  Intelligence, pp 203--209

\bibitem[{Yang et~al.(2015)Yang, Liu, Zhao, Sun, and Chang}]{yang2015network}
Yang C, Liu Z, Zhao D, Sun M, Chang EY (2015) Network representation learning
  with rich text information. In: Proceedings of the 24th International Joint
  Conference on Artificial Intelligence, pp 2111--2117

\bibitem[{Yang et~al.(2017)Yang, Wang, Li, Zhang, and Li}]{yang2017properties}
Yang D, Wang S, Li C, Zhang X, Li Z (2017) From properties to links: deep
  network embedding on incomplete graphs. In: Proceedings of the 2017 ACM on
  Conference on Information and Knowledge Management, ACM, pp 367--376

\bibitem[{Zhang et~al.(2016{\natexlab{a}})Zhang, Yin, Zhu, and
  Zhang}]{zhang2016collective}
Zhang D, Yin J, Zhu X, Zhang C (2016{\natexlab{a}}) Collective classification
  via discriminative matrix factorization on sparsely labeled networks. In:
  Proceedings of the 25th ACM International Conference on Information and
  Knowledge Management, ACM, pp 1563--1572

\bibitem[{Zhang et~al.(2016{\natexlab{b}})Zhang, Yin, Zhu, and
  Zhang}]{zhang2016homophily}
Zhang D, Yin J, Zhu X, Zhang C (2016{\natexlab{b}}) Homophily, structure, and
  content augmented network representation learning. In: Proceedings of the
  16th IEEE International Conference on Data Mining, IEEE, pp 609--618

\bibitem[{Zhang et~al.(2017)Zhang, Yin, Zhu, and Zhang}]{zhang2017user}
Zhang D, Yin J, Zhu X, Zhang C (2017) User profile preserving social network
  embedding. In: Proceedings of the 26th International Joint Conference on
  Artificial Intelligence, pp 3378--3384

\bibitem[{Zhang et~al.(2018)Zhang, Yin, Zhu, and Zhang}]{zhang2018network}
Zhang D, Yin J, Zhu X, Zhang C (2018) Network representation learning: A
  survey. IEEE Transactions on Big Data

\end{thebibliography}
%
%

\end{document}